\documentclass[dvips,???]{imsart}
\usepackage{graphicx}
\usepackage[numbers]{natbib}
\usepackage{amsmath,amsfonts,mathrsfs,bm,graphicx,subfigure,alg}
\usepackage{fullpage}
\usepackage[small,bf]{caption}
\usepackage{algorithmic}

\newcommand{\eop}{{\hfill\vbox{\hrule height .2pt
      \hbox{\vrule width.2pt height 6pt
      \kern 4pt
      \vrule width .2pt}
      \hrule height .2pt}} \par\bigskip}
\RequirePackage[OT1]{fontenc}
\RequirePackage{amsthm,amsmath}
\RequirePackage{natbib}

\startlocaldefs

\newtheorem{theorem}{Theorem}
\newtheorem{corollary}{Corollary}

\newtheorem{proposition}{Proposition}

\endlocaldefs

\begin{document}
\begin{frontmatter}

\title{Split Bregman method for large scale fused Lasso}
\runtitle{Split Bregman method for large scale fused Lasso}
\begin{aug}
\author{\fnms{Gui-Bo} \snm{Ye}
\ead[label=e1]{yeg@uci.edu}}
\and
\author{\fnms{Xiaohui} \snm{Xie}
\ead[label=e2]{xhx@ics.uci.edu}}
\address{Department of Computer Science, University of California, Irvine\\
Institute for Genomics and Bioinformatics, University of California, Irvine\\
\printead{e1,e2}}

\affiliation{University of California Irvine}
\end{aug}
\begin{abstract}
Ordering of regression or classification coefficients occurs in many real-world applications. Fused Lasso exploits this ordering by explicitly regularizing the differences between neighboring coefficients through an $\ell_1$ norm regularizer. However, due to nonseparability and nonsmoothness of the regularization term, solving the fused Lasso problem is computationally demanding. Existing solvers can only deal with problems of small or medium size, or a special case of the fused Lasso problem in which the predictor matrix is identity matrix. In this paper, we propose an iterative algorithm based on split Bregman method to solve a class of large-scale fused Lasso problems, including a generalized fused Lasso and a fused Lasso support vector classifier. We derive our algorithm using augmented Lagrangian method and prove its convergence properties. The performance of our method is tested on both artificial data and real-world applications including proteomic data from mass spectrometry and genomic data from array CGH. We demonstrate that our method is many times faster than the existing solvers, and show that it is especially efficient for large $p$, small $n$ problems.
\end{abstract}

\end{frontmatter}

\section{Introduction}\label{Sec Introduction}

Regularization terms that encourage sparsity in coefficients are increasingly being used in regression and classification procedures. One widely used example is the Lasso procedure for linear regression, which minimizes the usual sum of squared errors, but additionally penalizes the $\ell_1$ norm of the regression coefficients. Because of the non-differentiability  of the $\ell_1$ norm, the Lasso procedure tends to shrink the regression coefficients toward zero and achieves sparseness. Fast and efficient algorithms are available to solve Lasso with as many as millions of variables, which makes it an attractive choice for many large-scale real-world applications.

The fused Lasso method introduced by \citet{TSRZK:JRSS:2005} is an extension of Lasso, and considers the situation where there is certain natural ordering in  regression coefficients. Fused Lasso takes this natural ordering into account by placing an additional regularization term on the differences of ``neighboring" coefficients.  Consider the linear regression of $\{(\mathbf{x}_i,y_i)\}_{i=1}^n$, where $\mathbf{x}_i=(x_{i1},\ldots,x_{ip})^T$ are the predictor variables and $y_i$ are the responses. (We assume $x_{ij},y_i$ are standardized with zero mean and unit variance across different observations.) Fused Lasso finds the coefficients of linear regression by minimizing the following loss function

\begin{equation}\label{fused lasso two}
\Phi(\beta)=\frac{1}{2}\sum_{i=1}^n\left(y_i-\sum_{j=1}^px_{ij}\beta_j\right)^2+\lambda_1\sum_{i=1}^p|\beta_i|
+\lambda_2\sum_{i=2}^p|\beta_i-\beta_{i-1}|,
\end{equation}
where the regularization term with parameter $\lambda_1$ encourages the sparsity of the regression coefficients, while the regularization term with parameter $\lambda_2$ shrinks the differences between neighboring coefficients toward zero. As such, the method achieves both sparseness and smoothness in the regression coefficients.

Regression or classification variables with some inherent ordering occur naturally in many real-world applications.  In genomics, chromosomal features such as copy number variations (CNV), epigenetic modification patterns, and genes are ordered naturally by their chromosomal locations.  In proteomics, molecular fragments from mass spectrometry (MS) measurements are ordered by their mass-to-charge ratios (m/z). In dynamic gene network inference, gene regulatory networks from developmentally closer cell types are more similar than those from more distant cell types \citep{ahmed2009recovering}.  Fused Lasso exploits these natural ordering,  so, not surprisingly, it has found applications in these areas particularly suitable. For example,  Tibshirani and Wang successfully applied  fused Lasso to detect DNA copy number variations in tumor samples using array comparative genomic hybridization (CGH) data \citep{TW:Biostatisics:2008}.  Tibshirani et al. used fused Lasso to  select proteomic features that can separate tumor vs normal samples  \citep{TSRZK:JRSS:2005}.  In addition to the application areas mentioned above, fused Lasso or the extension of it has also found applications in a number of other areas, including image denoising,  social networks \citep{ahmed2009recovering}, quantitative trait network analysis \citep{kim2009multivariate}, etc.

The loss function in \eqref{fused lasso two} is strictly convex, so a global optimal solution is guaranteed to exist. However, finding the optimal solution is computationally challenging due to the nondifferentiability of $\Phi(\beta)$.  Existing methods circumvent the nondifferentiability of $\Phi(\beta)$ by introducing  $2p-1$ additional variables and converting the unconstrained optimization problem into a constrained one with $6p-1$ linear inequality constraints.  Standard convex optimization tools such as SQOPT  \citep{PMM:2006}  and CVX \citep{GBY:online:2010} can then be applied.  Because of the large number of variables introduced, these methods are computationally demanding in terms of both time and space, and, in practice, have only been able to solve fused Lasso problems with small or medium sizes.

Component-wise coordinate descent has been proposed as an efficient approach for solving many $l_1$ regularized convex optimization problems, including Lasso, grouped Lasso, elastic nets, graphical Lasso, logistic regression, etc \citep{FHHT:AAS:2007}.  However, coordinate descent cannot be applied to the fused Lasso problem because the variables in the loss function  $\Phi(\beta)$ are nonseparable due to the second regularization term, and as such, convergence is not guaranteed \citep{tseng2001convergence}.

For a special class of fused Lasso problems, named fused Lasso signal approximator (FLSA), where the predictor variables $x_{ij}=1$ for all $i=j$ and $0$ otherwise, there are algorithms available to solve it efficiently.  A key observation of FLSA first noticed by Friedman et al. is that for fixed $\lambda_1$, increasing $\lambda_2$ can only cause pairs of variables to fuse and they  become unfused for any larger values of $\lambda_2$.  This observation allows Friedman et al. to develop a fusion algorithm to solve FLSA for a path of $\lambda_2$ values by keeping track of fused variables and using coordinate descent for component-wise optimization.  The fusion algorithm was later extended and generalized by Hoefling \citep{Hoefling:Arxiv:2009}.  However, for the fusion algorithm to work, the solution path as a function of $\lambda_2$ has to be piecewise linear , which is not true for the general fused Lasso problem \citep{rosset2007piecewise}.  As such,  these algorithms are not applicable to the general fused Lasso case.

In this paper, we propose a new method based on the split Bregman iteration for solving the general fused Lasso problem.  Although the Bregman iteration was an old technique proposed in the sixties \citep{Bregman:Akad:1967,Cetin:SP:1989}, it gained significant interest only recently after  Osher and his coauthors demonstrated its high efficiency for image restoration   \citep{OBGXY:MMS:2005,GO:SIAM:2009,CaiOsherShen:MMS:2009}.  Most recently, it has also been shown to be an efficient tool for compressed sensing \citep{COS:MC:2009,OMDY:CMS:2010,YOGD:SIAM:2008}, matrix completion \citep{CCS:SIOPT:2010} and low rank matrix recovery \citep{CLMW:Arxive:2009}.  In the following, we will show that the general fused Lasso problem can be reformulated so that split Bregman iteration can be readily applied.

The rest of the paper is organized as follows. In Section \ref{sec algorithm}, we  derive algorithms for a class of fused Lasso problems from augmented Lagrangian function including SBFLasso for general fused Lasso, SBFLSA for FLSA and SBFLSVM for fused Lasso support vector classifier. The convergence properties of our algorithms are also presented. We demonstrate the performance and effectiveness of the algorithm through numerical examples in Section \ref{sec numerial experiments}, and describe additional implementation details. Algorithms described in this paper are implemented in Matlab and are freely available from the authors.

\section{Algorithms}\label{sec algorithm}

\subsection{Split Bregman iteration for a generalized fused Lasso problem}

We first describe our algorithm in a more general setting than the one described in \eqref{fused lasso two}.  Instead of the quadratic error function, we allow the error function to be any convex function of the regression coefficients.  In addition, we relax the assumption that the coefficients should be ordered along a line as in \eqref{fused lasso two}, and allow the ordering to be specified arbitrarily, e.g., according to a graph.  For the generalized fused Lasso, we find $\beta$ by solving the following unconstrained optimization problem
\begin{equation}\label{eq generalized fused lasso}
\min_{\beta} V(\beta) + \lambda_1 \| \beta \|_1 + \lambda_2 \| L \beta\|_1,
\end{equation}
where $V(\beta)=V(\beta; X,y)$ is the error term, the regularization term with parameter $\lambda_1$ encourages the sparsity of $\beta$ as before, and the regularization term with parameter $\lambda_2$ shrinks the differences between neighboring variables as specified in matrix $L$ toward zero.  We assume $L$ is  an $m\times p$ matrix.  In the standard fused Lasso in \eqref{fused lasso two}, $L$ is simply a $(p-1) \times p$ matrix with zeros entries everywhere except $1$ in the diagonal and $-1$ in the superdiagonal.  The unconstrained problem \eqref{eq generalized fused lasso} can be reformulated into an equivalent constrained problem
\begin{align}
\min_{\beta}\quad &V(\beta)+\lambda_1 \| a \|_1 + \lambda_2 \|b \|_1 \nonumber\\
{\rm s.t.} \quad &a=\beta \nonumber\\
& b= L\beta. \label{eq generalized fused lasso constrained}
\end{align}

Although split Bregman methods originated from Bregman iterations \citep{COS:MC:2009,OBGXY:MMS:2005,YOGD:SIAM:2008,ZBBO:SIIMS:2010,GO:SIAM:2009,CaiOsherShen:MMS:2009}, it is more convenient to derive the split Bregman method for the generalized Lasso using the augmented Lagrangian method \citep{Hestenes:JOTA:1969,Rockafellar:MP:1973}.

Note that the Lagrangian function of \eqref{eq generalized fused lasso constrained} is
\begin{equation}\label{eq lagrangian}
\tilde{\mathcal{L}}(\beta,a,b,u,v)=V(\beta)+\lambda_1\|a\|_1 + \lambda_2\|b\|_1+\langle u, \beta-a\rangle + \langle v, L\beta-b\rangle,
\end{equation}
where $u\in \mathbb{R}^p$ is a dual variable corresponding to the linear constraint $\beta=a$, $v\in \mathbb{R}^m$ is a dual variable corresponding to the linear constraint $L\beta=b$,
 $\langle \cdot,\cdot\rangle$ denotes the standard inner product in Euclidean space.
The augmented Lagrangian function of \eqref{eq generalized fused lasso constrained} is similar to the \eqref{eq lagrangian} except for adding two terms $\frac{\mu_1}{2}\|\beta-a\|_2^2 + \frac{\mu_2}{2}\|L\beta-b\|_2^2$ to penalize the violation of linear constraints $\beta=a$ and $L\beta=b$. That is,
\begin{align}
\mathcal{L}(\beta,a,b,u,v)=V(\beta)+\lambda_1\|a\|_1 + \lambda_2\|b\|_1+\langle u, \beta-a\rangle + \langle v, L\beta-b\rangle + 
 \frac{\mu_1}{2}\|\beta-a\|_2^2 + \frac{\mu_2}{2}\|L\beta-b\|_2^2,
\end{align}
where  $\mu_1>0$ and $\mu_2>0$ are two parameters.

Consider the problem of finding a saddle point $(\beta^*,a^*,b^*,u^*,v^*)$ of the augmented Lagrangian function $\mathcal{L}(\beta,a,b,u,v)$ such that
\begin{equation}\label{eq saddle point problem}
\mathcal{L}(\beta^*, a^*, b^*, u, v) \le \mathcal{L}(\beta^*, a^*, b^*, u^*, v^*) \le \mathcal{L}(\beta, a, b, u^*, v^*)
\end{equation}
for all $\beta$, $a$, $b$, $u$ and $v$. It can be shown that $\beta^*$ is an optimal solution of \eqref{eq generalized fused lasso} if and only if $(\beta^*,a^*,b^*,u^*,v^*)$ solves the above saddle point problem  for some $a^*$, $b^*$, $u^*$, and $v^*$.

 We solve the saddle point problem through an iterative algorithm by alternating between the primal and the dual optimization as follows
\begin{equation}\label{eq ALM primal dual}
\begin{cases}
{\rm Primal:} \quad (\beta^{k+1},a^{k+1},b^{k+1})=\arg\min_{\beta,a,b}\mathcal{L}(\beta,a,b,u^k,v^k) \\
{\rm Dual:}  \quad\quad u^{k+1}=u^k+\delta_1 (\beta^{k+1}-a^{k+1}),\ v^{k+1}=v^k+\delta_2 (L\beta^{k+1}-b^{k+1})
\end{cases}
\end{equation}
where the first step updates the primal variables based on the current estimate of $u^k$ and $v^k$, while the second step updates the dual variables based on the current estimate of the primal variables.  Since the augmented Lagrangian function  is linear in $u$ and $v$, updating the dual variables are relatively easy and we use gradient ascent algorithm with step size $\delta_1$ and $\delta_2$.

The efficiency of the iterative algorithm \eqref{eq ALM primal dual} lies on whether the primal problem can be solved quickly.  The augmented Lagrangian function $\mathcal{L}$ still contains  nondifferentiable terms. But different from the original objective function in \eqref{eq generalized fused lasso}, the $\ell_1$ induced nondifferentiability has now been transferred from terms involving $\beta$ to terms involving $a$ and $b$ only. Moreover, the nondifferentiable terms involving $a$ and $b$ are now completely decoupled, and thus we can solve the primal problem by alternating minimization of $\beta$, $a$ and $b$,
\begin{equation}\label{eq ALM primal}
\begin{cases}
\beta^{k+1}=\arg\min_\beta   V(\beta)+\langle u^k, \beta-a^k\rangle + \langle v^k, L\beta-b^k\rangle +
 \frac{\mu_1}{2}\|\beta-a^k\|_2^2 + \frac{\mu_2}{2}\|L\beta-b^k\|_2^2\\
a^{k+1}=\arg\min_a  \lambda_1\|a\|_1 + \langle u^k, \beta^{k+1}-a\rangle  + \frac{\mu_1}{2}\|\beta^{k+1}-a\|_2^2\\
b^{k+1}=\arg\min_b  \lambda_2\|b\|_1+  \langle v^k, L\beta^{k+1}-b\rangle + \frac{\mu_2}{2}\|L\beta^{k+1}-b\|_2^2.
\end{cases}
\end{equation}

Minimization of $a$ and $b$ in \eqref{eq ALM primal} can be done efficiently using soft thresholding, because the objective functions are quadratic and nondifferentiable terms are completely separable.
 Let $\mathcal{T}_\lambda$ be a soft thresholding operator defined on vector space and satisfying
\begin{equation}\label{eq threshold operator}
\mathcal{T}_{\lambda}(w)=[t_\lambda(w_1),t_\lambda(w_2),\ldots,\ldots]^T,\quad\hbox{with}\ t_\lambda(w_i)=\hbox{sgn}(w_i)\max\{0,|w_i|-\lambda\}.
\end{equation}
Using the soft thresholding operator, the optimal solution of $a$ and $b$ in \eqref{eq ALM primal} can be written as
\begin{equation}\label{eq iteration minmization two}
a^{k+1}=\mathcal{T}_{\mu_1^{-1}\lambda_1}(\beta^{k+1}+\mu_1^{-1}u^k)\quad\hbox{and}\quad
b^{k+1}=\mathcal{T}_{\mu_2^{-1}\lambda_2}(L\beta^{k+1}+\mu_2^{-1}v^k).
\end{equation}
Therefore, the efficiency of the iterative algorithm depends entirely on whether the minimization of $\beta$ in \eqref{eq ALM primal} can be done efficiently. If $V(\beta)$ is a quadratic function as in the standard fused Lasso, the optimal solution $\beta^{k+1}$ can be found analytically.

In theory, the alternate minimization between the primal variables needs to run multiple times until convergence. However, we do not have to completely solve the primal problem since it is only one step of the overall iterative algorithm. Our algorithm uses only one alternation. Overall, we propose Algorithm \ref{alg generalized fused Lasso} for solving the saddle point problem \eqref{eq saddle point problem}, and consequently the problem \eqref{eq generalized fused lasso}.

\begin{algorithm}[htp]
\caption{Split Bregman method for the generalized Fused Lasso \eqref{eq generalized fused lasso}}
\label{alg generalized fused Lasso}
\begin{algorithmic}
\STATE Initialize $\beta^0$, $a^0, b^0, u^0$, and $v^0$.
\REPEAT
\STATE 1) $\beta^{k+1}=\arg\min_\beta   V(\beta)+\langle u^k, \beta-a^k\rangle + \langle v^k, L\beta-b^k\rangle + \frac{\mu_1}{2}\|\beta-a^k\|_2^2 + \frac{\mu_2}{2}\|L\beta-b^k\|_2^2$
\STATE 2) $a^{k+1}=\mathcal{T}_{\mu_1^{-1}\lambda_1}(\beta^{k+1}+\mu_1^{-1}u^k)$
\STATE 3) $b^{k+1}=\mathcal{T}_{\mu_2^{-1}\lambda_2}(L\beta^{k+1}+\mu_2^{-1}v^k)$
\STATE 4) $u^{k+1}=u^k+\delta_1 (\beta^{k+1}-a^{k+1})$
\STATE 5) $v^{k+1}=v^k+\delta_2 (L\beta^{k+1}-b^{k+1})$
\UNTIL{\STATE Convergence}
\end{algorithmic}
\end{algorithm}

The convergence property of Algorithm \ref{alg generalized fused Lasso} is shown in the following theorem, which we prove in the Supplementary Info.
\begin{theorem}\label{theorem convergence analysis}
 Suppose there exists at least one solution $\beta^*$ of \eqref{eq generalized fused lasso}. Assume $V(\beta)$ is convex, $0<\delta\leq \mu_1$, $0<\delta_2\leq\mu_2$, and $\lambda_1>0$, $\lambda_2>0$. Then the following property for the split Bregman iteration in Algorithm \ref{alg generalized fused Lasso} holds:
\begin{equation}
\lim_{k\rightarrow\infty} V(\beta)+\lambda_1\|\beta^k\|_1+\lambda_2\|L\beta^k\|_1=V(\beta^*)+\lambda_1\|\beta^*\|_1+\lambda_2\|L\beta^*\|_1.
\end{equation}
Furthermore,
\begin{equation}
\lim_{k\rightarrow\infty}\|\beta^k-\beta^*\|=0
\end{equation}
whenever \eqref{eq generalized fused lasso} has a unique solution.
\end{theorem}

Note that the condition for the convergence in Theorem \ref{theorem convergence analysis} is quite easy to satisfy. $\lambda_1,\lambda_2$ are regularization parameters and should always be larger than zero. So as long as $0<\delta_1\leq \mu_1$ and $0<\delta_2\leq \mu_2$, the algorithm converges. In our implementation, we just choose $\delta_1=\mu_1$ and $\delta_2=\mu_2$.

\subsection{Split Bregman for the standard fused Lasso (SBFLasso)}
Next we apply Algorithm \ref{alg generalized fused Lasso} to solve the standard fused Lasso problem \eqref{fused lasso two}, which constitutes a special case of the generalized fused Lasso problem with
\begin{equation*}
V(\beta; X, y) = \frac{1}{2} \| X\beta -y\|_2^2\quad
\end{equation*}
and
\begin{equation}
L\beta=(\beta_2-\beta_1,\beta_3-\beta_2,\ldots,\beta_p-\beta_{p-1})^T \label{eq difference operator},
\end{equation}
where $X=(x_{i,j})_{i=1,j=1}^{n,p}$ and $y=(y_1,\ldots,y_n)^T$.

The objective function on minimizing $\beta$ in Algorithm \ref{alg generalized fused Lasso} is now quadratic and differentiable, and thus the optimal solution can be found by solving a set of linear equations:
\begin{equation}\label{eq linear equation SBFLasso}
(X^TX+\mu_1I+\mu_2L^TL)\beta^{k+1}=X^Ty+\mu_1(a^k-\mu_1^{-1}u^k)
+\mu_2L^T(b^k-\mu_2^{-1}v^k),
\end{equation}
while the other four steps in Algorithm \ref{alg generalized fused Lasso} are easy to implement and can be computed quickly. So the efficiency of the algorithm largely depends on how fast the linear equations can be solved. Matrix $D=X^TX+\mu_1I+\mu_2L^TL$ is a $p\times p$ matrix, independent of the optimization variables.  For small $p$, we can invert $D$ and store $D^{-1}$ in the memory, so the linear equations can be solved with minimal cost. However, for large $p$, we will need to numerically solve the linear equations at each iteration.

The matrix $P=\mu_1I+\mu_2L^TL$ occurring in \eqref{eq linear equation SBFLasso} is a tridiagonal positive definite matrix, and  as such, the linear equation $(\mu_1I+\mu_2L^TL)\mathbf{x}=\mathbf{g}$ can be solved efficiently for any $\mathbf{x},\mathbf{g}\in \mathbb{R}^p$, requiring only order of $p$ operations. More specifically, there exists a matrix $\hat{L}$ satisfying $\hat{L}_{ij}=0$ for all $i\neq j$ and $i \neq j-1$ such that $(\mu_1+1)I+\mu_2L^TL=\hat{L}\hat{L}^T$. This decomposition can be achieved by using Cholesky factorization. Thus solving equation \eqref{eq linear equation FLSA} is equivalent to solving two systems of linear equations $\hat{L}z=X^Ty+\mu_1(a^k-\mu_1^{-1}u^k)+\mu_2L^T(b^k-\mu_2^{-1}v^k)$ and $\hat{L}^T\beta^{k+1}=z$. These two equations can be easily solved due to the special structure of $\hat{L}$.

The linear system  \eqref{eq linear equation} is very special for large $p$, small $n$ problems in that $X^TX$ will be a low rank matrix with rank at most $n$. In combination of the special structure of matrix $P$ mentioned above, we use preconditioned conjugate gradient algorithm (PCG) to solve \eqref{eq linear equation}.

The PCG algorithm \citep{saad:book:2003} computes an approximate solution of the linear equations $H\mathbf{x}=\mathbf{g}$ using a preconditioner $P$, where both $H,P\in \mathbb{R}^{p\times p}$ are symmetric positive definite.  For the linear equation \eqref{eq linear equation SBFLasso}, we  use preconditioner $P=\mu_1I+\mu_2L^TL$ and the PCG algorithm converges in less than $n$ steps. In our numerical implementation, we found that PCG converges in a few steps much smaller than $n$.

\subsection{Split Bregman for FLSA (SBFLSA)}\label{subsec SBFLSA}

The fused Lasso signal approximator (FLSA) problem introduced by Friedman et al. corresponds to a special case of the standard fused Lasso with the predictor variables $X$ being an identify matrix. Therefore, for FLSA, the primal problem \label{eq linear equation} minimizing $\beta$ is simply
\begin{equation}\label{eq linear equation FLSA}
((\mu_1+1)I+\mu_2L^TL)\beta^{k+1}=y+\mu_1(a^k-\mu_1^{-1}u^k)+\mu_2L^T(b^k-\mu_2^{-1}v^k),
\end{equation}
while other iterations stay the same.

As mentioned in the preceding section, matrix $(\mu_1+1)I+\mu_2L^TL$ is a tridiagonal positive definite matrix, so \eqref{eq linear equation FLSA} can be solved very efficiently, requiring only order of $p$ operations. Therefore, we have a fast solver for SBFLSA using the split Bregman iteration in Algorithm \ref{alg generalized fused Lasso}.

\subsection{Iterative algorithm for fused Lasso Support Vector Classifier (SBFLSVM)}
Next we derive a split Bregman algorithm for the fused Lasso support vector classifier (FLSVM) introduced by \citep{TSRZK:JRSS:2005}. FLSVM uses a hinge loss function \citep{Vapnik:book:1998} for two-label classification problems. It finds the optimal classification coefficients $(\hat{\beta}_0,\hat{\beta})$ that minimize
\begin{equation}\label{eq fused lasso SVM}
f(\beta_0,\beta)=\frac{1}{n}\sum_{i=1}^n(1-y_i(\beta^T\mathbf{x}_i+\beta_0))_++\lambda_1\|\beta\|_1+\lambda_2\|L\beta\|_1,
\end{equation}
where $(u)_+=\max\{u,0\}$ for any $u\in \mathbb{R}$ and $L$ is the difference operator defined in \eqref{eq difference operator}.

Because the hinge loss function  $(1-t)_+$ is not differentiable, the primal problem involving $\beta$ is now more difficult to solve. As a result, we will not directly apply Algorithm \ref{alg generalized fused Lasso} to solve FLSVM. Instead, we introduce additional variables to deal with the nondifferentiability of the hinge loss.

Let $Y$ be a diagonal matrix with its diagonal elements to be the vector $y$. The unconstrained problem in \eqref{eq fused lasso SVM} can be reformulated into an equivalent constrained optimization problem
\begin{align}
\min_{\beta,\beta_0,a,b,c}\ & \frac{1}{n}\sum_{i=1}^n(c_i)_++\lambda_1\|a\|_1+\lambda_2\|b\|_1\nonumber\\
\ \ \ s.\ t.\quad & \beta=a\nonumber\\
\qquad\qquad & L\beta=b\nonumber\\
\qquad\qquad & \mathbf{1}-YX\beta-\beta_0y=c,\label{eq fused lasso SVM new}
\end{align}
where $\mathbf{1}$ is an $n$-column  vector of $1$s.

The augmented Lagrangian function of \eqref{eq fused lasso SVM new} is
\begin{align*}
\mathcal{L}(\beta,\beta_0,a,b,c,u,v,w)=\frac{1}{n}\sum_{i=1}^n(c_i)_++\lambda_1\|a\|_1+\lambda_2\|b\|_1+\langle u,\beta-a\rangle+\langle v,L\beta-b\rangle+\\
\langle w,\mathbf{1}-YX\beta-\beta_0-c\rangle
+\frac{\mu_1}{2}\|\beta-a\|_2^2+\frac{\mu_2}{2}\|L\beta-b\|_2^2+
\frac{\mu_3}{2}\|\mathbf{1}-YX\beta-\beta_0-c\|_2^2,
\end{align*}
where $u,v,w$ are dual variables corresponding to linear constraints $\beta=a$, $L\beta=b$, and $1-YX\beta-\beta_0y=c$ respectively.  Positive reals $\mu_1,\mu_2$ and $\mu_3$ are the penalty parameters for the violation of the linear constraints.

Similar to the derivation of Algorithm \ref{alg generalized fused Lasso}, we find the saddle point of $\mathcal{L}$ by iteratively updating the primal and dual directions
\begin{equation}\label{eq iteration minimization FLSVM one}
\begin{cases}
(\beta^{k+1},\beta_0^{k+1})=\arg\min_{\beta,\beta_0}\langle u^k,\beta-a^k\rangle
+\langle v^k,L\beta-b^k\rangle+\langle w^k,\mathbf{1}-YX\beta-\beta_0y-c^k\rangle\cr
\qquad\qquad\qquad\qquad + \frac{\mu_1}{2}\|\beta-a^k\|_2^2
+\frac{\mu_2}{2}\|L\beta-b^k\|_2^2+\frac{\mu_3}{2}\|\mathbf{1}-YX\beta-\beta_0y-c^k\|_2^2,\cr
a^{k+1}= \arg\min_a \lambda_1\|a\|_1+\langle u^k,\beta^{k+1}-a\rangle+\frac{\mu_1}{2}\|\beta^{k+1}-a\|_2^2,\cr
b^{k+1}=\arg\min_b\lambda_2\|b\|_1+\langle v^k,L\beta^{k+1}-b\rangle+\frac{\mu_2}{2}\|L\beta^{k+1}-b\|_2^2,\cr
c^{k+1}=\arg\min_c\frac{1}{n}\sum_{i=1}^n(c_i)_++\langle w^k,\mathbf{1}-YX\beta^{k+1}-\beta_0^{k+1}y-c\rangle\cr
\qquad\qquad\qquad\quad +\frac{\mu_3}{2}\|\mathbf{1}-YX\beta^{k+1}-\beta_0^{k+1}y-c\|_2^2\cr
u^{k+1}=u^k+\delta_1(\beta^{k+1}-a^{k+1}),\cr
v^{k+1}=v^k+\delta_2(L\beta^{k+1}-b^{k+1}),\cr
w^{k+1}=w^k+\delta_3(\mathbf{1}-YX\beta^{k+1}-\beta_0^{k+1}y-c^{k+1})
\end{cases}
\end{equation}
The update for $a^{k+1},b^{k+1},u^{k+1},v^{k+1},w^{k+1}$ are almost the same as the one in Algorithm \ref{alg generalized fused Lasso}, so we focus on the update for $(\beta^{k+1},\beta_0^{k+1})$ and $c^{k+1}$. In Supplementary Information we show that PCG can still be applied to solve the updating of $(\beta^{k+1},\beta_0^{k+1})$ with some modifications.

To update of $c^{k+1}$ in \ref{eq iteration minimization FLSVM one}, we use the following proposition, which is proven in the Supplementary Info.
\begin{proposition}\label{prop shrinkage operator}
Let $s_\lambda(w)=\arg\min_{x\in\mathbb{R}}\lambda x_++\frac{1}{2}\|x-w\|_2^2$. Then
\begin{equation}\label{eq svm shrinkage operator}
s_\lambda(w)=\left\{\begin{array}{ll}
w-\lambda,&w>\lambda,\\0,&0\leq w\leq \lambda,\\
w,&w<0.\end{array}\right.
\end{equation}
\end{proposition}

With Proposition \ref{prop shrinkage operator}, we can then update $c^{k+1}$ in \ref{eq iteration minimization FLSVM one} according to
\begin{corollary}\label{cor solution}
$c^{k+1}=\mathcal{S}_{\frac{1}{n\mu_3}}(\mathbf{1}-YX\beta^{k+1}-\beta_0^{k+1}y+\mu_3^{-1}w^k)$
is the solution of equation \eqref{eq svm shrinkage equation one}, where
$$\mathcal{S}_\lambda(w)=(s_\lambda(w_1),s_\lambda(w_2),\ldots,s_\lambda(w_n)),\quad\forall w\in \mathbb{R}^n$$ with $s_\lambda$ defined by \eqref{eq svm shrinkage operator}.
\end{corollary}
\begin{proof}
The equation \eqref{eq svm shrinkage equation one} is equivalent to
\begin{equation}\label{eq svm shrinkage equation two}
c^{k+1}=\arg\min_c\frac{1}{n\mu_3}\sum_{i=1}^n(c_i)_++\frac{1}{2}\|\mathbf{1}-YX\beta^{k+1}-\beta_0^{k+1}y-c+\mu_3^{-1}w^k\|_2^2.
\end{equation}
Note that each element of $c$ is independent of each other in \eqref{eq svm shrinkage equation two},  we can get the desired result by using Proposition \ref{prop shrinkage operator}.
\end{proof}

In summary, we derive Algorithm \ref{alg SBFLSVM} to solve \eqref{eq fused lasso SVM}.
\begin{algorithm}[htp]
\caption{Split Bregman method for FLSVM}
\label{alg SBFLSVM}
\begin{algorithmic}
\STATE Initialize $\beta^0, \beta_0^0$, $a^0$, $b^0$, $c^0$, $u^0$, $v^0$, and $w^0$.
\REPEAT
\STATE 1) Update $\beta^{k+1},\beta_0^{k+1}$ by solving the linear equations:
\STATE
$
\left(\begin{matrix}
\mu_1I+\mu_2L^TL+\mu_3X^TY^2X&\mu_3X^TYy\cr
\mu_3y^TYX&\mu_3y^Ty
\end{matrix}\right)\left(\begin{matrix}\beta^{k+1}\cr\beta_0^{k+1}\end{matrix}\right)$
\STATE
$=\mu_1\left(\begin{matrix}a^k-\mu_1^{-1}u^k\cr 0\end{matrix}\right)+\mu_2\left(\begin{matrix}L^T\cr 0\end{matrix}\right)(b^k-\mu_2^{-1}v^k)+
\mu_3\left(\begin{matrix}X^TY\cr y^T\end{matrix}\right)(\mathbf{1}-c^k+\mu_3^{-1}w^k)
$
\STATE 2) $a^{k+1}=\mathcal{T}_{\mu_1^{-1}\lambda_1}(\beta^{k+1}+\mu_1^{-1}u^k)$
\STATE 3) $b^{k+1}=\mathcal{T}_{\mu_2^{-1}\lambda_2}(L\beta^{k+1}+\mu_2^{-1}v^k)$
\STATE 4) $c^{k+1}=\mathcal{S}_{\frac{1}{n\mu_3}}(\mathbf{1}-YX\beta^{k+1}-\beta_0^{k+1}y+\mu_3^{-1}w^k)$
\STATE 5) $u^{k+1}=u^k+\delta_1 (L\beta^{k+1}-a^{k+1})$
\STATE 6) $v^{k+1}=v^k+\delta_2 (\beta^{k+1}-v^{k+1})$
\STATE 7) $w^{k+1}=w^k+\delta_3(\mathbf{1}-YX\beta^{k+1}-\beta_0^{k+1}y-c^{k+1})$
\UNTIL{\STATE Convergence}
\end{algorithmic}
\end{algorithm}

The convergence property of Algorithm \ref{alg SBFLSVM} is shown in the following theorem, which we prove in the Supplementary Info.
\begin{theorem}\label{theorem convergence analysis SVM}
Suppose there exists at least one solution $\beta^*$ of \eqref{eq fused lasso SVM}. Assume $0<\delta_1\leq \mu_1, 0<\delta_2\leq \mu_2$ and $\lambda_1>0,\lambda_2>0$. Then the following property for
Algorithm \ref{alg SBFLSVM} holds:
\begin{align}\label{eq SBFLSVM convergence energy}
\lim_{k\rightarrow\infty}\ \frac{1}{n}\sum_{i=1}^n(1-y_i(\mathbf{x}_i^T\beta^k+\beta_0^k))_++\lambda_1\|\beta^k\|_1+\lambda_2\|L\beta^k\|_1
\nonumber\\ =\frac{1}{n}\sum_{i=1}^n(1-y_i(\mathbf{x}_i^T\beta^*+\beta_0^*))_+
+\lambda_1\|\beta^*\|_1+\lambda_2\|L\beta^*\|_1.
\end{align}
Furthermore,
\begin{equation}\label{eq SBFLSVM convergence beta}
\lim_{k\rightarrow\infty}\|\beta^k-\beta^*\|=0
\end{equation}
whenever \eqref{eq fused lasso SVM} has a unique solution.
\end{theorem}

\section{Experimental Results}\label{sec numerial experiments}
Next we illustrate the efficiency of split Bregman method for fused Lasso using time trials on artificial data as well as real-world applications from genomics and proteomics. All our algorithms were implemented in Matlab, and compiled on a windows platform. Time trials were generated on an Intel Core 2 Duo desktop PC (E7500, 2.93GHz).

As the regression form of the fused Lasso procedures is more frequently used, we will thus focus on testing the performance of SBFLasso and SBFLSA. To evaluate the performance of SBFLasso, we compare it with SQOPT and CVX. SQOPT \citep{PMM:2006} is used in the original fused Lasso paper by Tibshirani et al. \citep{TSRZK:JRSS:2005}. It is a two-phase active set algorithm, designed for quadratic programming problems with sparse linear constraints. CVX is a general convex optimization package  \citep{GBY:online:2010}. SQOPT and CVX solve the fused Lasso by introducing additional variables and constraints to transform the nondifferentiable objective function into a smooth one. Both solvers  are implemented in Matlab, and thus are directly comparable to our implementation. SQOPT allows warm start, so we will use it whenever possible. To evaluate the performance of SBFLSA for solving FLSA, we mainly compare it with the path algorithm proposed by Hoefling \citep{Hoefling:Arxiv:2009}.

The stopping criterion of SBFLasso is specified as follows. Let $\Phi(\beta^k)=\frac{1}{2}\|X\beta^k-y\|_2^2+\lambda_1\|\beta^k\|_1+\lambda_2\|L\beta^k\|_1$.
According to Theorem \ref{theorem convergence analysis}, $\lim_{k\rightarrow \infty}\Phi(\beta^k)=\Phi(\beta^*)$. Therefore, we terminate SBFLasso when the relative change of the energy functional $\frac{1}{2}\|X\beta-y\|_2^2+\lambda_1\|\beta\|_1+\lambda_2\|L\beta\|_1$  falls below certain threshold $\delta$. We used $\delta=10^{-5}$ in our simulation, i.e., we stop the Bregman iteration whenever
\begin{equation}
RelE:=\frac{|\Phi(\beta^{k+1})-\Phi(\beta^k)|}{\Phi(\beta^k)}\leq 10^{-5}.\label{eq relative energy}
\end{equation}

Note that the convergence of Algorithm \ref{alg generalized fused Lasso} is guaranteed no matter what values of $\mu_1$ and $\mu_2$ are used as shown in Theorem \ref{theorem convergence analysis}. The speed of the algorithm can, however, be influenced by the choices of $\mu_1$ and $\mu_2$ as it would affect the number of iterations involved. In our implementation, we choose the parameter values using a pretrial procedure in which we test the convergence rate for a set of parameter values and identify the one that gives rise to the highest convergence rate. For regression problems, we always set $\mu_1=\mu_2$ and select values from the set $\{0.2,0.4,0.6,0.8,1\}\times \|y\|_2$. The parameter selecting procedure can certainly be further improved, but empirically we find it works well for all the problems we tested.

\subsection{Artificial data}
\subsubsection{Solving fused Lasso}\label{subsubsec solving fused lasso}
\begin{figure}[htp]
\subfigure[]{\includegraphics[width=.48\textwidth]{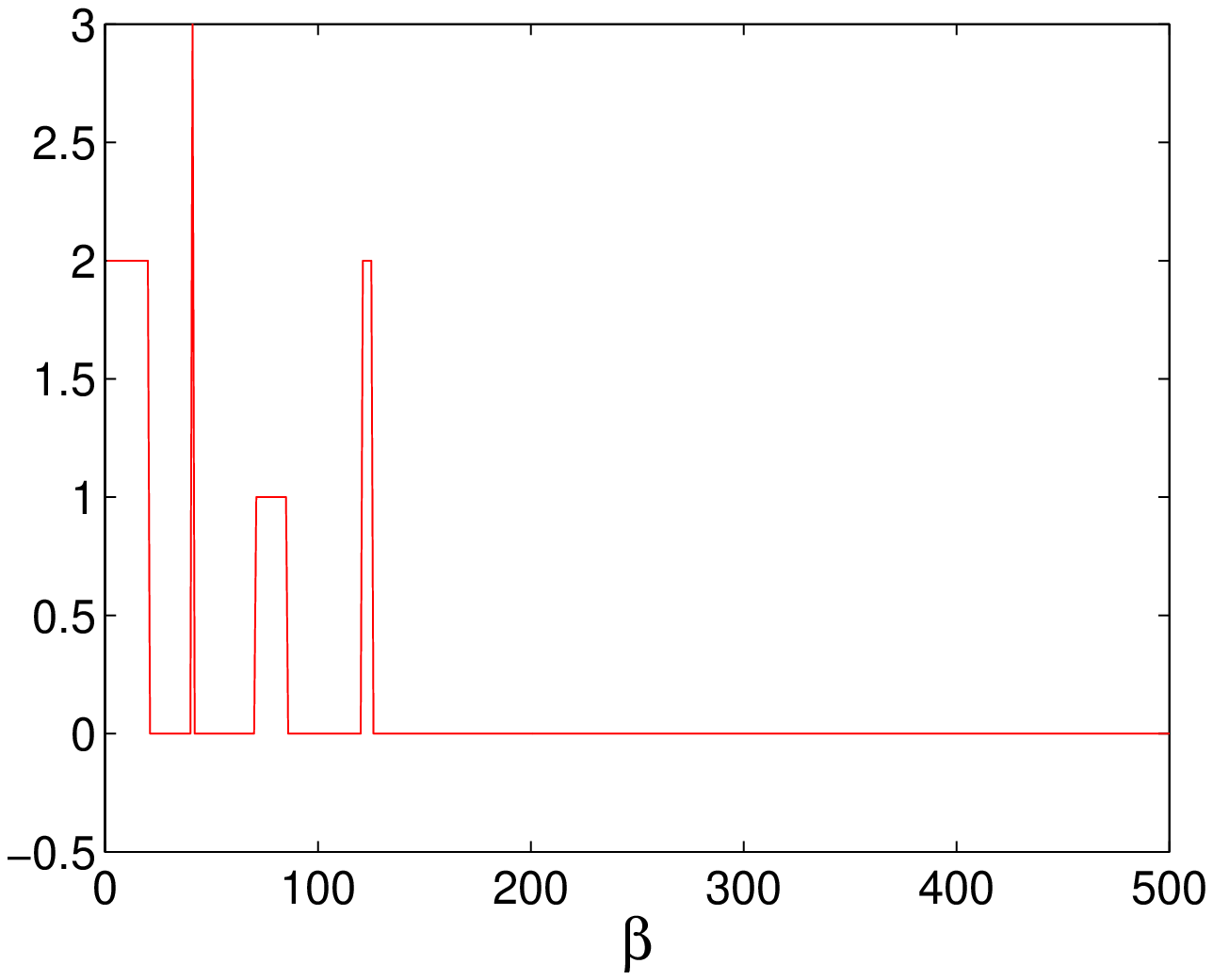}}
\subfigure[]{\includegraphics[width=.48\textwidth]{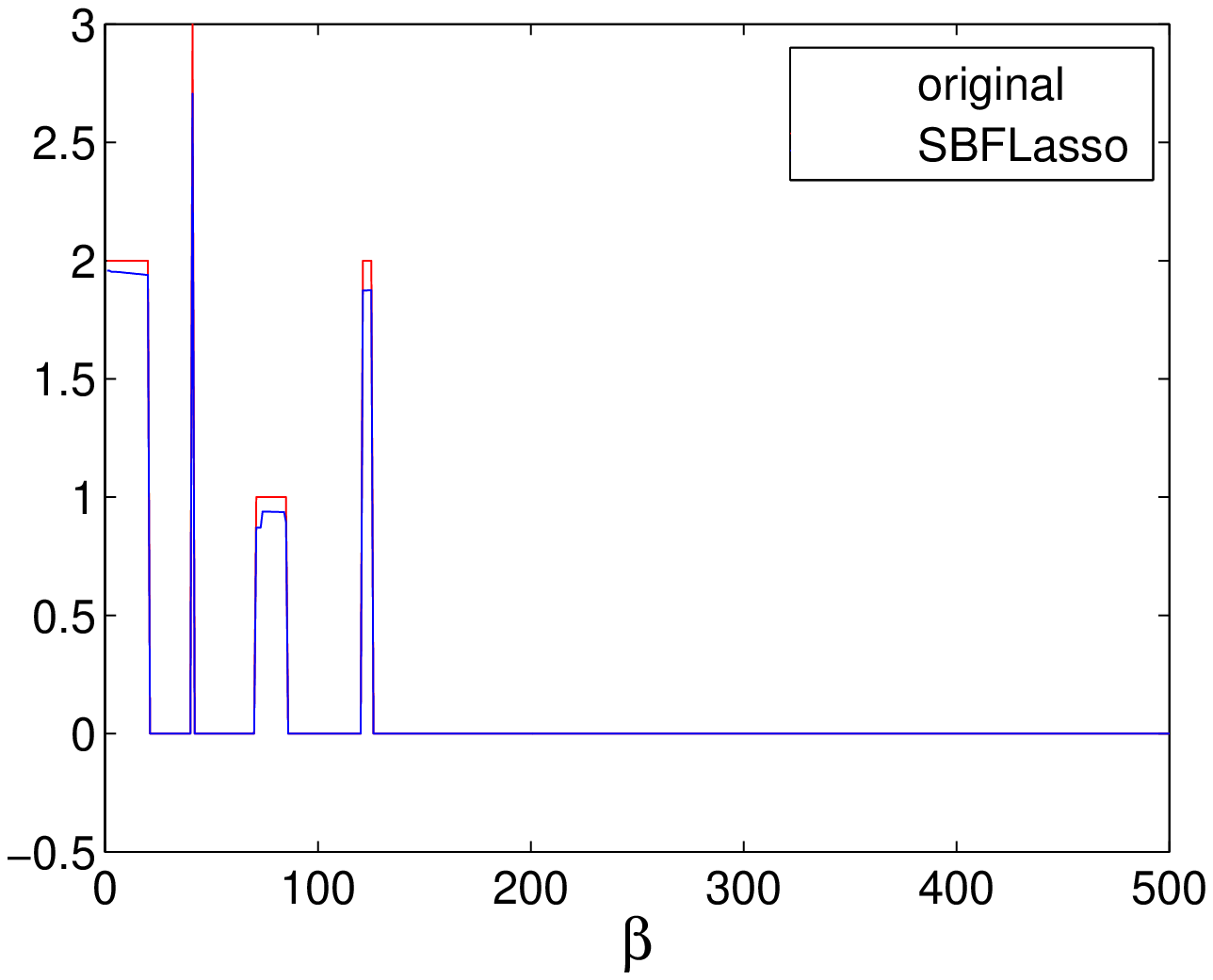}}
\caption{(a) The figure of coefficient $\beta$ in 500 dimension; (b) The blue line is the solution derived by SBFLasso with $\lambda_1=16,\lambda_2=20$ and the red line is the original $\beta$ in 500 dimension.}\label{figure one}
\end{figure}

We generated Gaussian data with $n$ observations and $p$ predictors, with each pair of predictors $X_i,X_j (i\neq j)$ having the same population correlation $\rho$.  The outcome values were generated by
$Y=\sum_{j=1}^p\beta_jX_j+\epsilon$,
where $\epsilon$ is the Gaussian noise with mean $0$ and variance $\sigma$. The regression coefficient  $\beta=(\beta_1,\ldots,\beta_p)$ is a sparse vector with the values of $\beta_j$ are generated according to
\begin{equation*}
\beta_i=\left\{\begin{array}{ll}2,&i=1,2\ldots,20,121,122,\ldots,125\\
3,&i=41\\
1,&i=71,72,\ldots,85\\
0,&\hbox{else}.
\end{array}\right.
\end{equation*}
\noindent The design of $\beta$ is motivated by the fact that fused lasso are especially suitable for coefficients that are constant for an interval and change in jumps. An example plot of $\beta$ of size $p=500$ can be seen in Figure \ref{figure one}(a).

Table \ref{table result one} shows the average CPU time for CVX, SQOPT and SBFLasso to solve the fused Lasso problems. SBFLasso consistently outperforms CVX in all cases we tested with a speedup of at least ten fold. Note that CVX fails to obtain results for large $p$ problems due to out of memory errors. Although it has similar performance to SQOPT for small size problems ($p\sim 200$), SBFLasso is significantly faster than SQOPT for large $p$ problems. For problems of $n=200,p=20000,$ SBFLasso is able to obtain the optimal solutions within $\sim 30$ seconds, while it takes about $800$ seconds for SQOPT to obtain the similar results. Overall, our algorithm is about twenty times faster than SQOPT for the large $p$ problems.

\begin{table}[htp]
\caption{ Run times (CPU seconds) for fused lasso problems of various sizes $p$ and $n$, different correlation $\rho$ between the features. Methods are SQOPT, CVX and SBFLasso. The results are averaged over $100$ runs (using $4$ different predictor matrix $X$ and $25$ different values of regularization parameters $\lambda_1,\lambda_2$ (or $s_1,s_2$).}
{\begin{tabular*}{0.9\textwidth}{@{\extracolsep{\fill}}lccccccc}
\hline &&n=100&n=100&n=200&n=200&n=200&n=200\\
 $\rho$&Method&p=200&p=1000&p=2000 &p=5000  &p=10000&p=20000\\
 &&time(sec)&time(sec)&time(sec)  &time(sec)&time(sec)&time(sec)\\
\hline  &CVX
&0.496 &2.646& 18.148 & 64.453&-&-\\
$\rho=0$&SQOPT  &0.0334& 0.510& 5.738& 39.269&147.534&$>600$\\
&SBFLasso &0.0366&0.155&1.488& 5.845&12.724&28.441\\
\hline &CVX
& 0.523 &2.792&16.812& 61.914&-&-\\
$\rho=0.2
$&SQOPT  &0.0323& 0.572& 6.812& 47.196&205.365&$>600$\\
&SBFLasso &0.0352&0.323&2.831&9.716&18.249&34.061\\\hline
&CVX
& 0.518 & 2.719& 16.504 &63.456&-&-\\
$\rho=0.4
$&SQOPT  &0.0299& 0.611&6.063&48.010&203.973&$>600$\\
&SBFLasso &0.0338&0.265&2.803&8.897&24.680&26.990\\\hline
&CVX
&0.510& 2.856 & 17.020 &62.920&-&-\\
$\rho=0.6
$&SQOPT  &0.0312&0.519& 6.508& 45.339&197.794&$>600$\\
&SBFLasso &0.0286&0.143&2.190& 8.947&20.586&36.157\\\hline
&CVX
&0.511&2.995 &19.379& 68.425&-&-\\
$\rho=0.8
$&SQOPT  &0.0293& 0.527& 5.678& 41.147&178.208&$>600$\\
&SBFLasso &0.0190&0.221&1.426& 6.446&15.505&41.614\\\hline
\end{tabular*}}
\label{table result one}
\end{table}

To evaluate how the performance of SBFLasso scales with problem size, we plotted the CPU time that SBFLasso took to solve the fused Lasso problem as a function of $p$ and $n$. Figure 2 shows such a curve, where CPU time is averaged over $500$ runs with different parameters $\lambda_1,\lambda_2$ and different design matrix $X$. We note that the CPU times are roughly linear in both $n$ and  $p$.

A key to the success of SBFLasso is that we split the regularization terms $\|\beta\|_1$ and $\|L\beta\|_1$ and make the minimization problems separable. Due to the soft thresholding in
the Bregman iteration, the solutions obtained by SBFLasso are naturally sparse as we
can see from Figure  \ref{figure one}(b). This is contrast to solutions obtained by CVX and SQOPT,
because no thresholding steps are involved, solutions obtained by these two algorithms
are not sparse, and sparseness can only be achieved through a thresholding step in the
postprocessing.

\begin{figure}[htp]
\subfigure[]{\includegraphics[width=.48\textwidth]{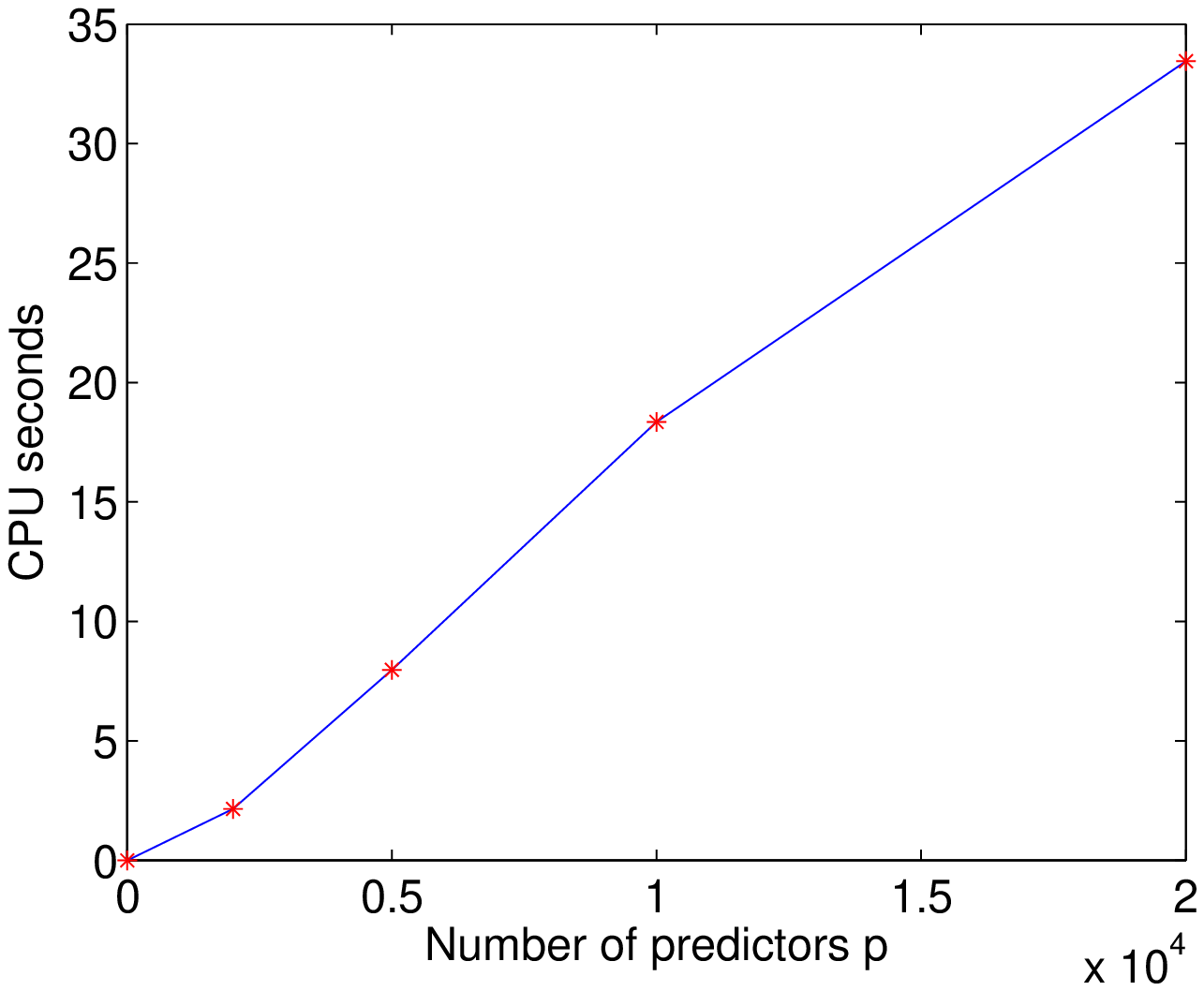}}
\subfigure[]{\includegraphics[width=.48\textwidth]{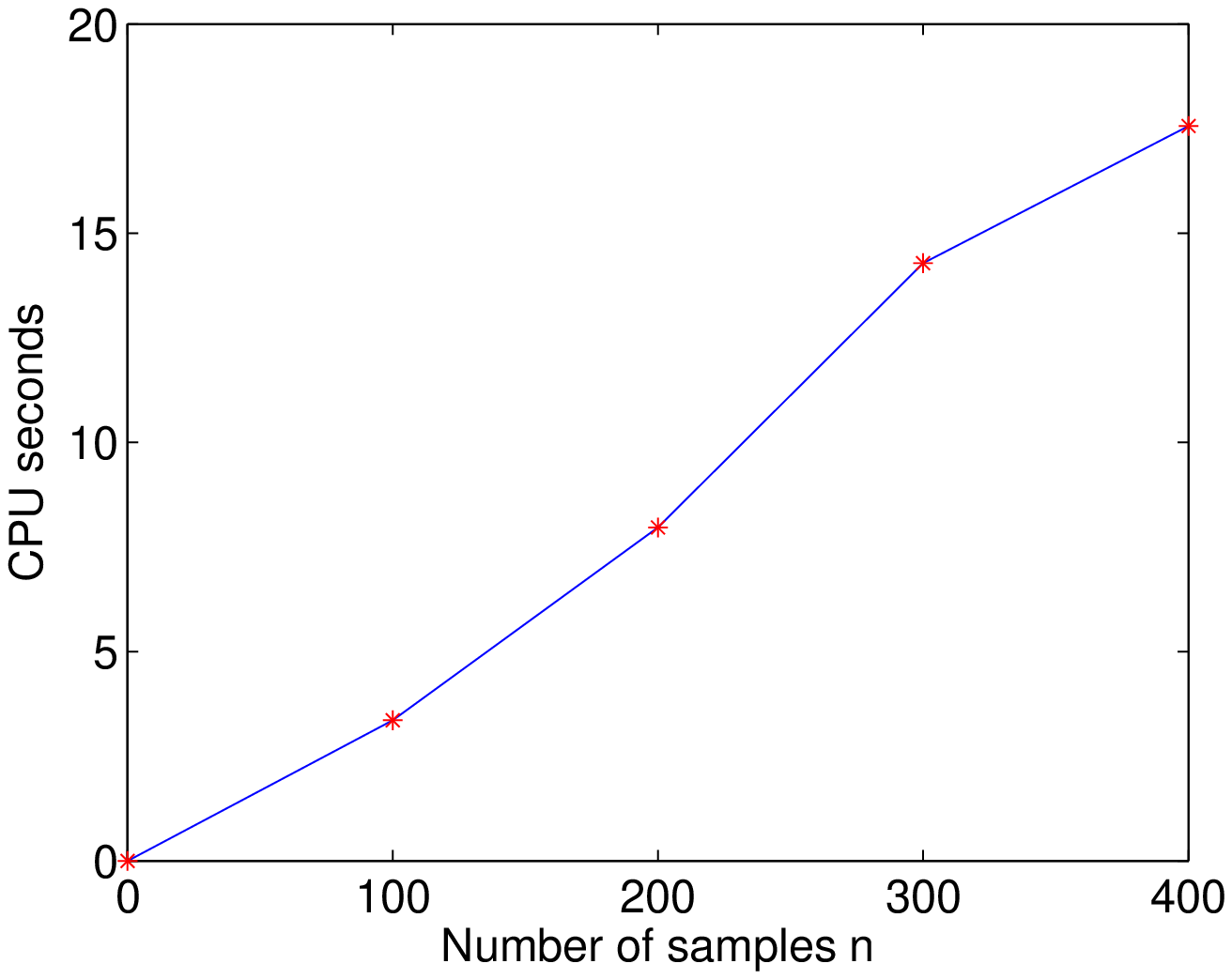}}
\caption{CPU times for SBFLasso for the same problem as in Table \ref{table result one}, for different values of $n$ and $p$. In each case the times are averaged over 500 runs. (a) $n$ is fixed and equals to $200$; (b) $p$ is fixed and equals to $5000$.}\label{figure CPU times}
\end{figure}

\subsubsection{Solving FLSA}
Next we compare SBFLSA and the path algorithm (PATHFLSA)
\citep{Hoefling:Arxiv:2009} for solving FLSA. PATHFLSA uses a fusion algorithm to solve FLSA,
taking advantage of the special structure of the error term. It represents the state of the
art for solving the FLSA problem. We generate data according to
$y=\beta+\epsilon$,
where $\epsilon$ is the Gaussian noise with mean $0$ and variance $\sigma$, and $\beta$ is a sparse vector which has similar shape as the one shown in Figure \ref{figure one}(a) with more nonzero entries. We vary $p$ from $10^3$ to $10^6$ and the results for each $p$ are averaged over $10$ runs.

Table \ref{table result two} shows that the computational times of SBFLSA and PATHFLSA for solving
the FLSA problems.We note that the performance of SBFLSA is similar to PATHFLSA
in almost all cases we tested, and both algorithms significantly outperforms SQOPT with
thousands of times faster for large p problems.

\begin{table}[htp]
\caption{Run times (CPU seconds) for an 1-dimensional FLSA problems of various sizes $p$. Methods are SQOPT, SBFLSA
and path algorithm for FLSA(PATHFLSA). The results are averaged over $10$ runs.}
{\begin{tabular*}{0.9\textwidth}{@{\extracolsep{\fill}}lcccc}
\hline
 parameters&Method&$p=10^4$&$p=10^5$&$p=10^6$\\
\hline  &SQOPT
&106.97 &$>5$\ hours&- \\
$\lambda_1=0.1,\lambda_2=0.8$&SBFLSA  &0.053& 0.754& 8.681\\
&PATHFLSA&0.050&0.651&8.685\\
\hline &SQOPT
&107.74&$>5$\ hours&-\\
$\lambda_1=0.2,\lambda_2=1.0
$&SBFLSA  &0.053& 0.820&8.263\\
&PATHFLSA &0.051&0.653&8.678\\\hline
&SQOPT
& 108.12 &$>5$\ hours& -\\
$\lambda_1=0.3,\lambda_2=1.2
$&SBFLSA &0.053& 0.798&9.289\\
&PATHFLSA &0.049&0.654&8.657\\\hline
&SQOPT
&106.13&$>5$\ hours& - \\
$\lambda_1=0.4,\lambda_2=1.5
$&SBFLSA &0.053&0.806& 9.892\\
&PATHFLSA &0.049&0.651&8.661\\
\hline
\end{tabular*}}
\label{table result two}
\end{table}

Although the performance of SBFLSA and PATHFLSA are similar for fixed $\lambda_1$ and
$\lambda_2$, PATHFLSA has an additional advantage of generating solutions for a path of the
regularization parameters. However, because PATHFLSA works by fusing variables, a
necessary condition for it to work is that the solution path has to be piece-wise linear
when varying $\lambda_2$. This condition is in general not true for both the fused Lasso and the
generalized fused Lasso. As such, it cannot be applied to these cases.

\subsection{Mass spectrometry data}
Mass spectrometry (MS) holds great promise for biomarker identification, and genome wide metabolic and proteomic profiling. The protein mass spectroscopy application was used as a motivating example for fused Lasso in the paper by  \citet{TSRZK:JRSS:2005}. Next we illustrate the efficiency of SBFLasso for solving the fused Lasso problem for mass spectrometry data. The data we use is taken from \cite{CDF:BMC:2009}. It consists of MS measurements of $95$ normal samples and $121$ samples taken from patients with ovarian cancer. The raw data contains a total of   $368,750$ mass-to-charge ratio (m/z) sites.

\begin{figure}[htp]
\subfigure[]{\includegraphics[width=.48\textwidth]{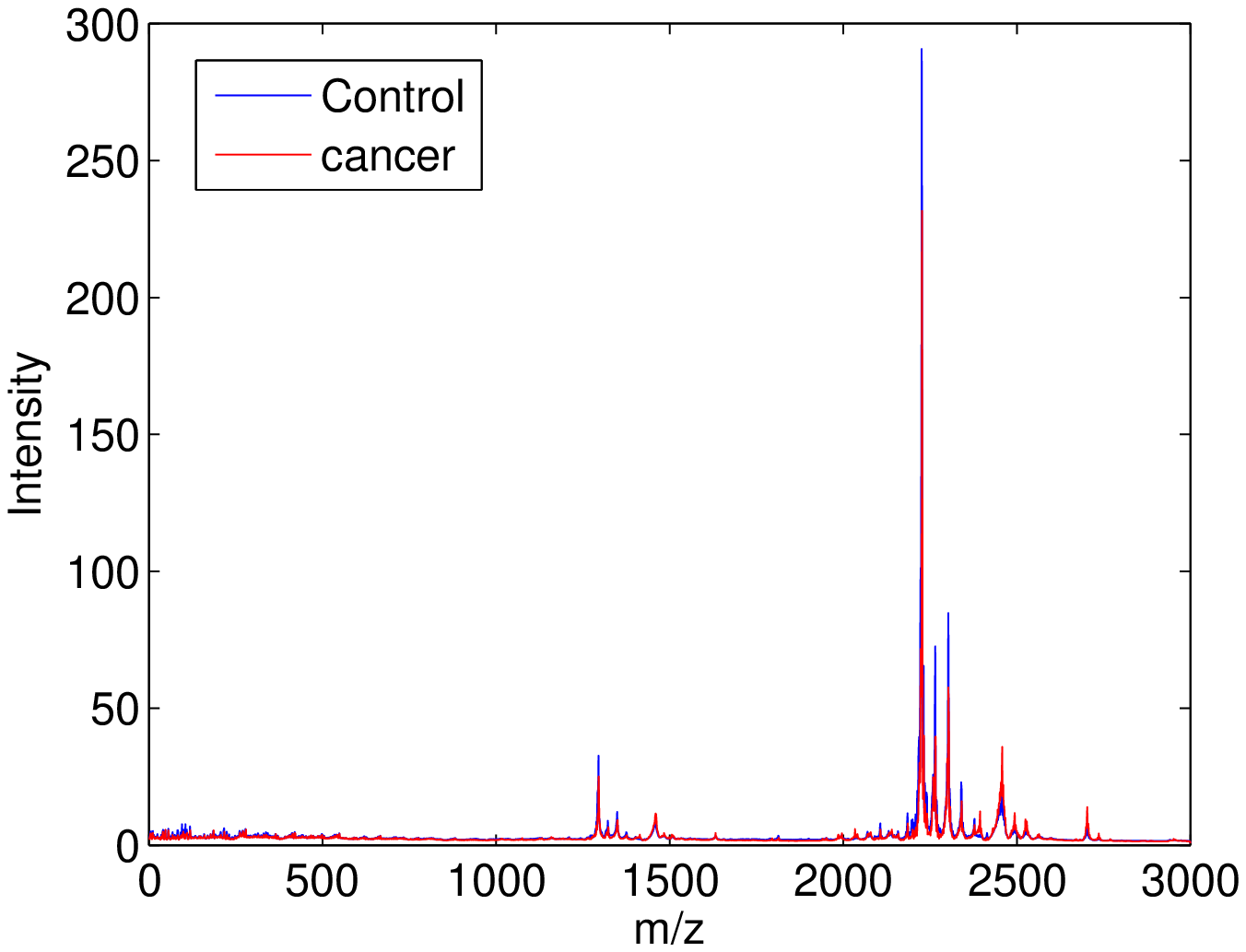}}
\subfigure[]{\includegraphics[width=.48\textwidth]{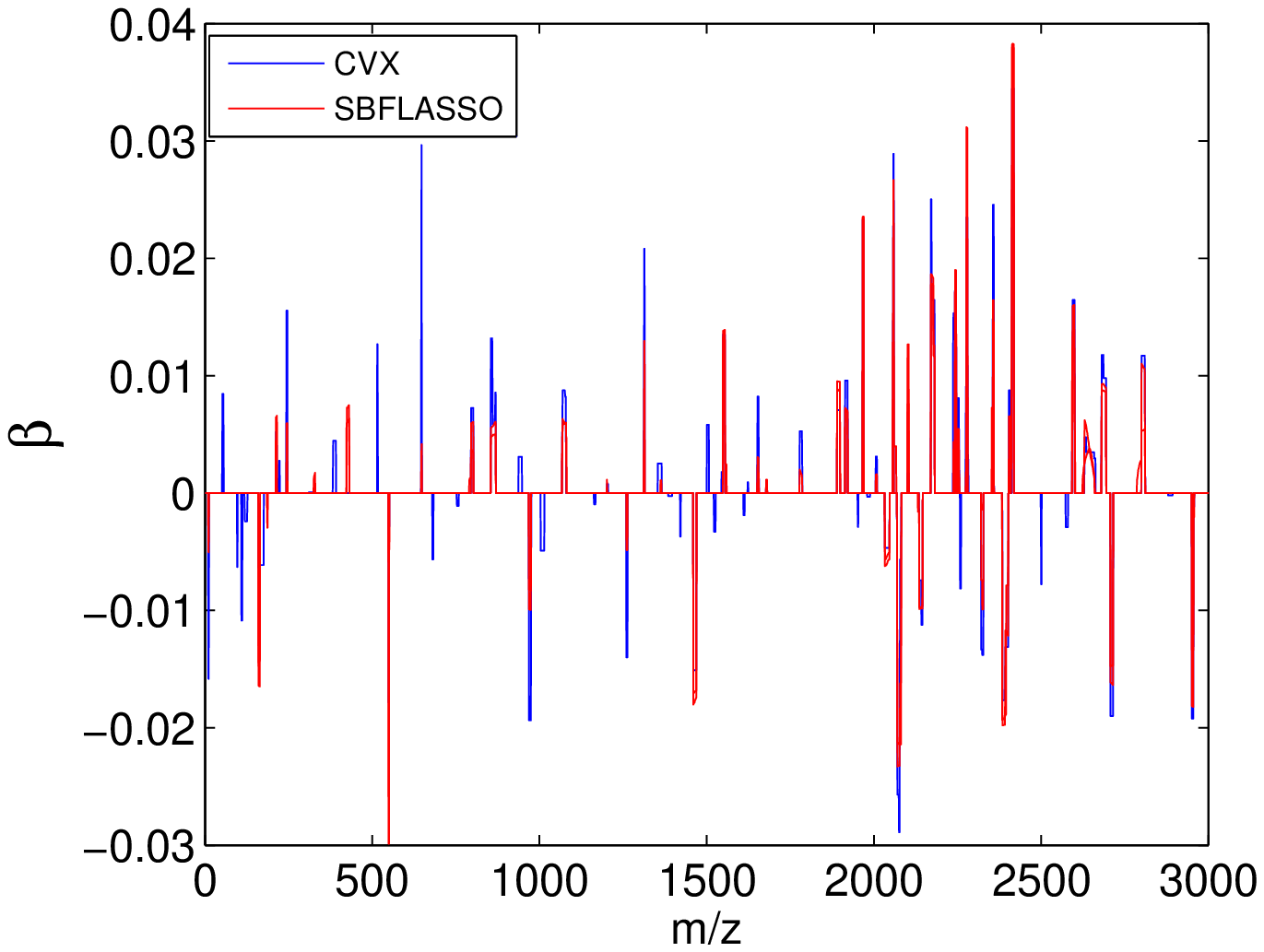}}
\caption{(a) Protein mass spectroscopy data:average profiles from normal (blue) and  cancer patients (red);
(b) Estimated $\beta$ from protein mass spectroscopy data with $\lambda_1=2,\lambda_2=3.5$ by CVX (blue) and SBFLasso (red).}\label{figure MS data}
\end{figure}

We first preprocessed the data using the procedure described in \citep{CDF:BMC:2009}, consisting of the following three steps: 1) re-sampling: Gaussian kernel reconstruction of the signal in order to have a set of $d$-dimensional vectors with equally spaced mass/charge values; 2) baseline correction: removes systematic artifacts, usually attributed to clusters of ionized matrix molecules hitting the detector during early portions of the experiment, or to detector overload; 3) normalization: corrects for differences in the total amount of protein desorbed and ionized from the sample plate.
The average profiles  from normal and cancer patients after preprocessing are shown in Figure \ref{figure MS data}.

\begin{table}[htp]
\caption{Run times (CPU seconds) for SBFLasso, SQOPT and CVX on MS data for different values of the regularization parameters $\lambda_1$ and $\lambda_2$.}
{\begin{tabular*}{0.75\textwidth}{@{\extracolsep{\fill}}lcccc}
\hline parameters& 10-CV error&SBFLasso&SQOPT&CVX\\\hline
 $\lambda_1=2.0,\lambda_2=3.5$&6/216&2.9854&31.707&24.481\\\hline
 $\lambda_1=2.5,\lambda_2=4.5$&8/216&3.6612&30.310&23.456\\\hline
 $\lambda_1=3.0,\lambda_2=1.0$&6/216&3.2082&35.911&22.261\\\hline
$\lambda_1=3.5,\lambda_2=2.5$&9/216&3.5080&32.094&21.130\\
 \hline
\end{tabular*}}
\label{table MS}
\end{table}

We apply the fused Lasso \eqref {fused lasso two} to the MS data to select features (m/z sites) that can be used to predict sample labels. For each sample, the response variable is either $1$ or $-1$, and the predictor variable is a vector consisting of the intensity of $p=3000$ sampled m/z sites. We used SBFLasso to solve the fused Lasso problem, and compared its performance to CVX and SQOPT. The results are summarized in Table \ref{table MS}, which shows the computational times spent by different solvers in a ten-fold cross-validation procedure for different parameters $\lambda_1$ and $\lambda_2$. SBFLasso is consistently many times faster than CVX and SQOPT, with an approximately ten-fold speedup in almost all cases. The coefficients derived by SBFLasso and the other solvers are very similar (Figure \ref{figure MS data}b), but SBFLasso is able to achieve a sparser solution, and in addition a slightly lower objective function than CVX.

\subsection{Comparative genomic hybridization (CGH) data}\label{subsection CGH}
In tumor cells, mutations often cause a large DNA segment to be deleted or inserted in a chromosome, in a phenomena called copy number variation (CNV).  Array CGH is a technique that is used to detect CNVs in a genome by labeling DNA from a test sample and normal reference sample differently using fluorophores and hybridizing to genomewide probes. The $\log$ ratio of the fluorescence intensity of the test DNA to the reference DNA is then calculated for each probe. A value greater than zero indicates a possible gain in DNA copies of the region around the probe, while a value less than zero suggests a possible loss. \cite{TW:Biostatisics:2008} demonstrated the efficiency of the fused lasso signal approximator (FLSA) for detecting CVNs using array CGH data. Next, we will show that SBFLSA is an efficient tool for solving the FLSA problem for array CGH data.

\begin{table}[htp]
\caption{Run times (CPU seconds) for SBFLSA on CGH data for different values of the regularization parameters $\lambda_1$ and $\lambda_2$.}
{\begin{tabular*}{0.75\textwidth}{@{\extracolsep{\fill}}lcccc}
\hline parameters& SBFLSA&PATHFLSA&SQOPT&CVX \\\hline
 $\lambda_1=0.10,\lambda_2=3.0$&0.007&0.006&0.7273&0.7124\\\hline
 $\lambda_1=0.12,\lambda_2=3.5$&0.007&0.005&0.6193&0.6451\\\hline
 $\lambda_1=0.15,\lambda_2=3.0$&0.007&0.006&0.6161&0.6604\\\hline
 $\lambda_1=0.18,\lambda_2=3.2$&0.006&0.005&0.6370&0.6244\\
 \hline
\end{tabular*}}
\label{table CGH}
\end{table}

\begin{figure}[htp]
\subfigure[]{\includegraphics[width=.48\textwidth]{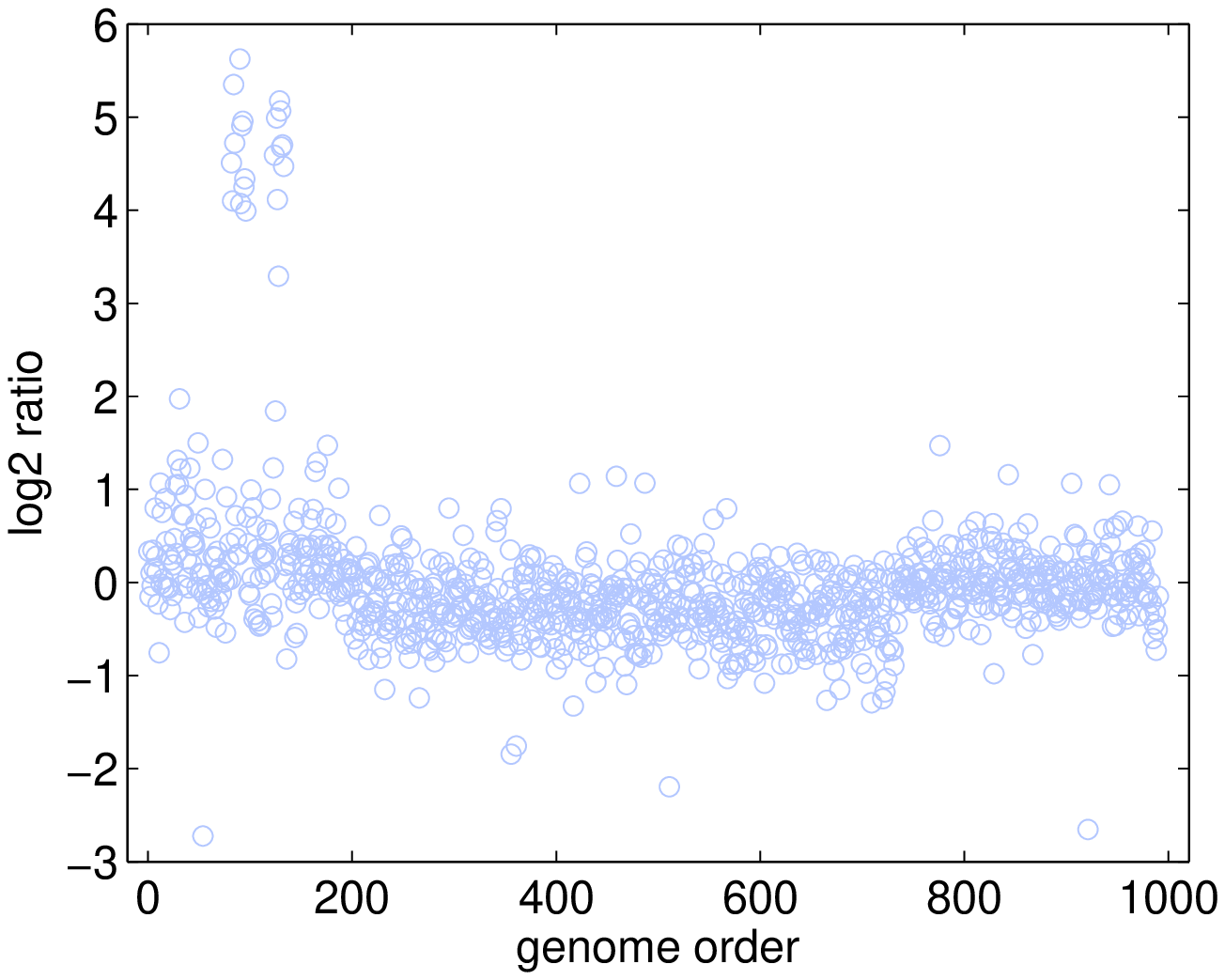}}
\subfigure[]{\includegraphics[width=.48\textwidth]{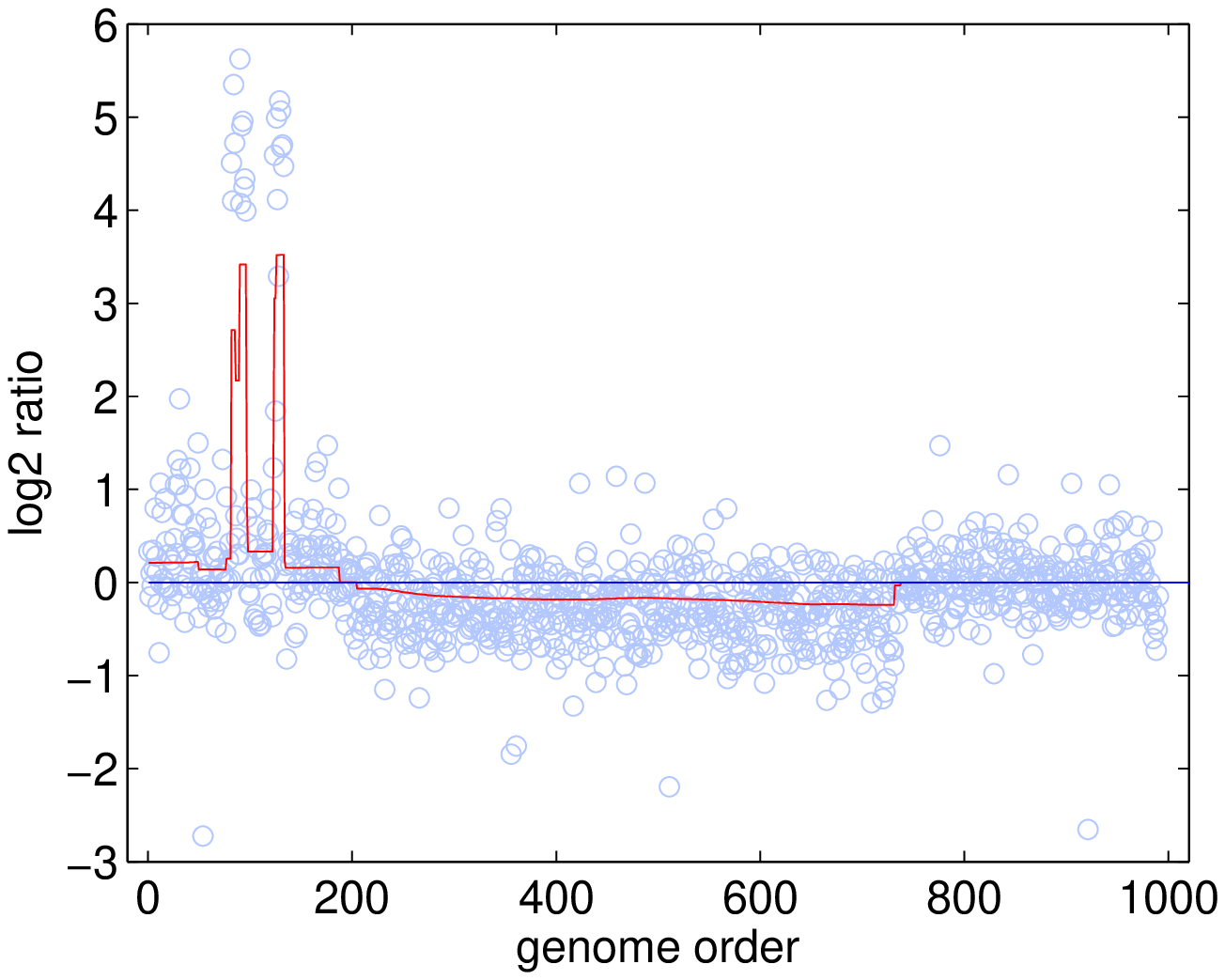}}
\caption{Fused lasso applied to some GBM data. The data are shown in the left panel, and the solid red line in the right panel represents the inferred copy  number $\hat{\beta}$ from SBFLSA. The gray line is for $y=0$.}\label{figure CGH data}
\end{figure}

We used the glioblastoma multiforme (GBM) data from \cite{BBJHVRS:CR:2005}, which contains array CGN profiling of samples from primary GBMs, a particular malignant type of brain tumor.  Table \ref{table CGH} shows the CPU times spent by SBFLSA to solve the FLSA problem for different regularization parameters $\lambda_1$ and $\lambda_2$. We observe that our method is significantly faster than SQOPT and CVX, with a speed improvement of about $100$ times. The performance of SBFLSA is also comparable to the path algorithm, which is specially designed, the state of the art for solving FLSA problems. Figure \ref{figure CGH data}(b) plots the copy number variants detected by SBFLSA with $\lambda_1=0.10$ and $\lambda_2=3.5$, clearly showing the gain of DNA segments in two nearby chromosomal regions in GBM.

\section{Discussion}

Fused Lasso is an attractive framework for regression or classification problems with some natural ordering occurring in regression or classification coefficients. It exploits this naturally ordering by explicitly regularizing the differences between neighboring coefficients through an $l_1$ norm regularizer. Solving the fused Lasso problem is, however, challenging because of the nondifferentiability of the objective function and the nonseparability of the variables involved in the nondifferentiable terms of the objective function. Existing solvers circumvent these difficulties by reformulating the fused Lasso problem into a smooth constrained convex optimization problem by introducing a large number of auxiliary variables and constraints, and as such, are only able to  solve small or medium size fused Lasso problems.

We derived an iterative algorithm based on the split Bregman method to solve a class of fused Lasso problems, including SBFLasso for the standard fused Lasso, SBFLSA for the fused Lasso signal approximator, and SBFLSVM for fused Lasso support vector classifier, and proved their convergence properties. Preliminary experimental results for SBFLasso and SBFLSA show their efficiency for large scale problems, especially for problems with large $p$, small $n$, which occur in many real-world applications.

The iterative algorithm we propose is very easy to implement, involving only a few lines of code. It is also very general and can be adapted to solve a variety of fused Lasso problems with minor modifications as we have shown.  In this aspect, it is very different from the path algorithm, which is specially designed for FLSA and requires significant amount of domain specific knowledge.  Because of its simplicity and generality, we expect the split Bregman iterative algorithm would find its usage in a wide range of other $\ell_1$ related regularization problems.

\bibliographystyle{unsrtnat}
\bibliography{ygb}

\newpage
\appendix
\section*{Appendix}
\subsection{Convergence analysis of Algorithm \ref{alg generalized fused Lasso}}
We use some similar ideas from \citet{CaiOsherShen:MMS:2009} to prove Theorem \ref{theorem convergence analysis}. Different from the case of \citet{CaiOsherShen:MMS:2009},  we do not require $V(\beta)$ to be differentiable, and treat the nondifferentiability of $V(\beta)$ explicitly by using its subgradient vector $h\in \partial V(\beta)$, see the proof for details.

{\bf Proof of Theorem \ref{theorem convergence analysis}:} Since all the subproblems involved in \eqref{eq ALM primal} are convex, the first order optimality condition of the Algorithm \ref{alg generalized fused Lasso} gives
\begin{equation}\label{eq optimality condition iteration GFLasso}
\begin{cases}
0=h^{k+1}+u^k+L^Tv^k+\mu_1(\beta^{k+1}-a^k)+\mu_2L^T(L\beta^{k+1}-b^k),\cr
\lambda_1p^{k+1}-u^k + \mu_1(a^{k+1}-\beta^{k+1})=0,\cr
\lambda_1q^{k+1}-v^k+\mu_2(b^{k+1}-L\beta^{k+1})=0,\cr
u^{k+1}=u^k+\delta_1(\beta^{k+1}-a^{k+1}),\cr
v^{k+1}=v^k+\delta_2(L\beta^{k+1}-b^{k+1}),
\end{cases}
\end{equation}
where $h^{k+1}\in \partial V(\beta^{k+1}), p^{k+1}\in \partial \|a^{k+1}\|_1$ and  $q^{k+1}\in \partial \|b^{k+1}\|_1$.

Since $\beta^*$ is a solution of \eqref{eq generalized fused lasso}, by the first order optimality condition, there exist $h^*,p^*,q^*$ such that
\begin{equation}\label{eq optimality condition GFLasso}
h^*+\lambda_1p^*+\lambda_2L^Tq^*=0
\end{equation}
where $h^*\in \partial V(\beta), p^*\in \partial \|\beta^*\|_1, q^*\in \partial \|b^*\|_1$ with $b^*=L\beta^*$. Introducing new variables $a^*=\beta^*, u^*=\lambda_1p^*,v^*=\lambda_2q^*$, we can formulate \eqref{eq optimality condition GFLasso} as
\begin{equation}\label{eq stationary optimality condition GFLasso}
  \begin{cases}
    0=h^*+u^*+L^Tv^*+\mu_1(\beta^*-a^*)+\mu_2L^T(L\beta^*-b^*),\quad \hbox{with}\quad h^*\in \partial V(\beta)\cr
\lambda_1p^*-u^* + \mu_1(a^*-\beta^*)=0,\quad \hbox{with}\quad  p^*\in \partial \|a^*\|_1,\cr
\lambda_1q^*-v^*+\mu_2(b^*-L\beta^*)=0,\quad \hbox{with}\quad q^*\in \partial \|b^*\|_1,\cr
u^*=u^*+\delta_1(\beta^*-a^*),\cr
v^*=v^*+\delta_2(L\beta^*-b^*).
  \end{cases}
\end{equation}
Comparing \eqref{eq stationary optimality condition GFLasso} with \eqref{eq optimality condition iteration GFLasso}, we can see that
$\beta^*,a^*,b^*,u^*,v^*$ is a fix point of Algorithm \ref{alg generalized fused Lasso}. Denote the errors by
$$\beta^k_e=\beta^k-\beta^*, a_e^k=a^k-a^*,b_e^k=b^k-b^*,u_e^k=u^k-u^*\ \hbox{and}\ v_e^k=v^k-v^*.$$
Subtracting the first equation of \eqref{eq optimality condition iteration GFLasso} by the first equation of \eqref{eq stationary optimality condition GFLasso}, we obtain
\begin{equation*}
  0=h^{k+1}-h^*+u_e^k+L^Tv_e^k+\mu_1(\beta_e^{k+1}-a_e^k)+\mu_2L^T(L\beta_e^{k+1}-b_e^k).
\end{equation*}
Taking the inner product of the left and right hand sides with respect to $\beta_e^{k+1}$, we have
\begin{eqnarray}
  0&=&\langle h^{k+1}-h^*,\beta^{k+1}-\beta^*\rangle+\langle u_e^k,\beta_e^{k+1}\rangle+\langle v_e^k,L\beta_e^{k+1}\rangle\nonumber\\
 &&+\mu_1\|\beta_e^{k+1}\|_2^2-\mu_1\langle a_e^k,\beta_e^{k+1}\rangle+\mu_2\|L\beta_e^{k+1}\|_2^2-\mu_2\langle b_e^k,L\beta_e^{k+1}\rangle
 \label{eq proof one}
\end{eqnarray}
Similarly, we can get
\begin{equation}
\lambda_1\langle p^{k+1}-p^*,a^{k+1}-a^*\rangle+\mu_1\|a_e^{k+1}\|_2^2-\mu_1\langle a_e^{k+1},\beta_e^{k+1}\rangle-\langle u_e^k,a_e^{k+1}\rangle=0,\label{eq proof two}
\end{equation}
\begin{equation}
\lambda_2\langle q^{k+1}-q^*,b^{k+1}-b^*\rangle+\mu_2\|b_e^{k+1}\|_2^2-\mu_2\langle b_e^{k+1},L\beta_e^{k+1}\rangle-\langle v_e^k,b_e^{k+1}\rangle=0.\label{eq proof three}
\end{equation}
Summing \eqref{eq proof one}, \eqref{eq proof two} and \eqref{eq proof three} together, we get
\begin{eqnarray}
 && \langle h^{k+1}-h^*,\beta^{k+1}-\beta^*\rangle +\lambda_1\langle p^{k+1}-p^*,a^{k+1}-a^*\rangle+\lambda_2\langle q^{k+1}-q^*,b^{k+1}-b^*\rangle\nonumber\\&&
 +\mu_1(\|\beta_e^{k+1}\|_2^2+\|a_e^{k+1}\|_2^2 -\langle \beta_e^{k+1},a_e^k+a_e^{k+1}\rangle) +\mu_2(\|L\beta_e^{k+1}\|_2^2+\|b_e^{k+1}\|_2^2
  \nonumber\\&&-\langle L\beta_e^{k+1},b_e^k+b_e^{k+1}\rangle) +\langle u_e^k,\beta_e^{k+1}-a_e^{k+1}\rangle+\langle v_e^k,L\beta_e^{k+1}-b_e^{k+1}\rangle=0\label{eq proof four}
\end{eqnarray}
Furthermore, by subtracting the fourth equation of \eqref{eq optimality condition iteration GFLasso} by the one of \eqref{eq stationary optimality condition GFLasso}, we have
\begin{equation*}
  u_e^{k+1}=u_e^k+\delta_1(\beta_e^{k+1}-a_e^{k+1}).
\end{equation*}
Taking square of both sides of the above equation implies
\begin{equation}
\langle u_e^k,\beta_e^{k+1}-a_e^{k+1}\rangle=\frac{1}{2\delta_1}(\|u_e^{k+1}\|_2^2-\|u_e^k\|_2^2)-
\frac{\delta_1}{2}\|\beta_e^{k+1}-a_e^{k+1}\|_2^2.\label{eq proof five}
\end{equation}
Similarly, we have
\begin{equation}
\langle v_e^k,L\beta_e^{k+1}-b_e^{k+1}\rangle=\frac{1}{2\delta_2}(\|v_e^{k+1}\|_2^2-\|v_e^k\|_2^2)-
\frac{\delta_2}{2}\|L\beta_e^{k+1}-b_e^{k+1}\|_2^2.\label{eq proof six}
\end{equation}
Substituting \eqref{eq proof five} and \eqref{eq proof six} into \eqref{eq proof four} yields
\begin{eqnarray}
&&\frac{1}{2\delta_1}(\|u_e^k\|_2^2-\|u_e^{k+1}\|_2^2)+\frac{1}{2\delta_2}(\|v_e^k\|_2^2-\|v_e^{k+1}\|_2^2)\nonumber\\
&=&\langle h^{k+1}-h^*,\beta^{k+1}-\beta^*\rangle +\lambda_1\langle p^{k+1}-p^*,a^{k+1}-a^*\rangle+\lambda_2\langle q^{k+1}-q^*,b^{k+1}-b^*\rangle\nonumber\\&& +\mu_1(\|\beta_e^{k+1}\|_2^2+\|a_e^{k+1}\|_2^2
-\langle \beta_e^{k+1},a_e^k+a_e^{k+1}\rangle-\frac{\delta_1}{2\mu_1}\|\beta_e^{k+1}-a_e^{k+1}\|_2^2)\nonumber
\end{eqnarray}
\begin{equation}
 +\mu_2(\|L\beta_e^{k+1}\|_2^2+\|b_e^{k+1}\|_2^2-\langle L\beta_e^{k+1},b_e^k+b_e^{k+1}\rangle-\frac{\delta_2}{2\mu_2}\|L\beta_e^{k+1}-b_e^{k+1}\|_2^2)\label{eq proof seven}
\end{equation}
Note that for any $\mathbf{x},\mathbf{y},\mathbf{z}\in \mathbb{R}^p$, we have
\begin{equation}
\|\mathbf{x}\|_2^2\pm\langle \mathbf{x},\mathbf{y}+\mathbf{z}\rangle+\|\mathbf{y}\|_2^2=\frac{1}{2}\|\mathbf{x}\pm \mathbf{y}\|_2^2
+\frac{1}{2}\|\mathbf{x}\pm \mathbf{z}\|_2^2+\frac{1}{2}(\|\mathbf{y}\|_2^2-\|\mathbf{z}\|_2^2).\label{eq elementary equation}
\end{equation}
Using this elementary equation, \eqref{eq proof seven} can be transformed to

\begin{eqnarray}
&&\frac{1}{2\delta_1}(\|u_e^k\|_2^2-\|u_e^{k+1}\|_2^2)+\frac{1}{2\delta_2}(\|v_e^k\|_2^2-\|v_e^{k+1}\|_2^2)\nonumber\\
&=&\langle h^{k+1}-h^*,\beta^{k+1}-\beta^*\rangle +\lambda_1\langle p^{k+1}-p^*,a^{k+1}-a^*\rangle+\lambda_2\langle q^{k+1}-q^*,b^{k+1}-b^*\rangle\nonumber\\&& +\frac{\mu_1}{2}\left(\|\beta_e^{k+1}-a_e^k\|_2^2+\|a_e^{k+1}\|_2^2
-\|a_e^k\|_2^2
+\frac{\mu_1-\delta_1}{\mu_1}\|\beta_e^{k+1}-a_e^{k+1}\|_2^2\right)\nonumber\\
&&+\frac{\mu_2}{2}\left(\|L\beta_e^{k+1}-b_e^k\|_2^2+\|b_e^{k+1}\|_2^2-\|b_e^k\|_2^2+\frac{\mu_2-\delta_2}{\mu_2}\|L\beta_e^{k+1}-b_e^{k+1}\|_2^2\right)
\nonumber
\end{eqnarray}
Summing the above equation from $k=0$ to $k=K$ yields
\begin{eqnarray}
 &&\frac{1}{2\delta_1}(\|u_e^0\|_2^2-\|u_e^{K+1}\|_2^2)+\frac{1}{2\delta_2}(\|v_e^0\|_2^2-\|v_e^{K+1}\|_2^2)
 \nonumber\\&&+\frac{\mu_1}{2}(\|a_e^0\|_2^2-\|a_e^{K+1}\|_2^2)+\frac{\mu_2}{2}(\|b_e^0\|_2^2-\|b_e^{K+1}\|_2^2)\nonumber\\
 &=&\sum_{k=0}^K\langle h^{k+1}-h^*,\beta^{k+1}-\beta^*\rangle+\lambda_1\sum_{k=0}^K \langle p^{k+1}-p^*,a^{k+1}-a^*\rangle
 \nonumber\\
 &&+\lambda_2\sum_{k=0}^K\langle q^{k+1}-q^*,b^{k+1}-b^*\rangle+\frac{\mu_1}{2}\sum_{k=0}^K\|\beta_e^{k+1}-a_e^k\|_2^2+\frac{\mu_1-\delta_1}{\mu_1}\sum_{k=0}^K\|\beta_e^{k+1}-a_e^{k+1}\|_2^2
 \nonumber\\&&+\frac{\mu_2}{2}\sum_{k=0}^K\|L\beta_e^{k+1}-b_e^k\|_2^2+\frac{\mu_2-\delta_2}{\mu_2}\sum_{k=0}^K\|L\beta_e^{k+1}-b_e^{k+1}\|_2^2\label{eq proof eight}
\end{eqnarray}
 The fact that $h^{k+1}\in \partial V(\beta^{k+1}),h^*\in \partial V(\beta^*)$ and $V(\beta)$ is convex implies
 \begin{eqnarray}
 &&\langle h^{k+1}-h^*,\beta^{k+1}-\beta^*\rangle \nonumber\\&=&V(\beta^{k+1})-V(\beta^*)-\langle h^*,\beta^{k+1}-\beta^*\rangle+V(\beta^*)-V(\beta^{k+1})
 -\langle h^{k+1},\beta^{k+1}-\beta^*\rangle\nonumber\\&\geq& 0\label{eq Bregman property}
 \end{eqnarray}
by the definition of subgradient.
Similarly, $\langle p^{k+1}-p^*,a^{k+1}-a^*\rangle\geq 0,\langle q^{k+1}-q^*,b^{k+1}-b^*\rangle\geq 0$. Together with the fact that $0<\delta_1\leq \mu_1$ and $0<\delta_2\leq \mu_2$, all terms involved in \eqref{eq proof eight} are  nonnegative. Therefore,
\begin{equation*}
  \sum_{k=0}^\infty\langle h^{k+1}-h^*,\beta^{k+1}-\beta^*\rangle\leq \frac{1}{2\delta_1}\|u_e^0\|_2^2+\frac{1}{2\delta_2}\|v_e^0\|_2^2
  +\frac{\mu_1}{2}\|a_e^0\|_2^2+\frac{\mu_2}{2}\|b_e^0\|_2^2
\end{equation*}
which leads to
$$\lim_{k\rightarrow \infty}\langle h^k-h^*,\beta^k-\beta^*\rangle=0.$$
Together with \eqref{eq Bregman property} leads to
\begin{equation}
\lim_{k\rightarrow \infty}V(\beta^k)-V(\beta^*)-\langle h^*,\beta^k-\beta^*\rangle=0.\label{eq proof nine}
\end{equation}
Similarly, we can prove
\begin{equation}
\lambda_1\lim_{k\rightarrow\infty} \|a^k\|_1-\|a^*\|_1-\langle a^k-a^*,p^*\rangle=0,\label{eq proof ten}
\end{equation}
\begin{equation}
\lambda_2\lim_{k\rightarrow\infty} \|b^k\|_1-\|b^*\|_1-\langle b^k-b^*,q^*\rangle=0,\label{eq proof eleven}
\end{equation}
\begin{equation}
  \lim_{k\rightarrow\infty}\|\beta^k-a^k\|_2=0\quad\hbox{and}\quad \lim_{k\rightarrow\infty}\|L\beta^k-b^k\|_2=0.\label{eq proof twelve}
\end{equation}
Since $\|\cdot\|_1$ is continuous, by \eqref{eq proof ten},\eqref{eq proof eleven} and \eqref{eq proof twelve}, we obtain
\begin{equation}
\lambda_1\lim_{k\rightarrow\infty} \|\beta^k\|_1-\|\beta^*\|_1-\langle \beta^k-\beta^*,p^*\rangle=0,\label{eq proof thirteen}
\end{equation}
\begin{equation}
\lambda_2\lim_{k\rightarrow\infty} \|L\beta^k\|_1-\|L\beta^*\|_1-\langle L\beta^k-L\beta^*,q^*\rangle=0.\label{eq proof fourteen}
\end{equation}
Summing \eqref{eq proof nine},\eqref{eq proof thirteen} and \eqref{eq proof fourteen} yields
\begin{eqnarray}
&&\lim_{k\rightarrow\infty}V(\beta^k)+\lambda_1\|\beta^k\|_1+\lambda_2\|L\beta^k\|_1-(V(\beta^*)+\lambda_1\|\beta^*\|_1+\lambda_2\|L\beta^*\|_1)
\nonumber\\&&\qquad+\langle \beta^k-\beta^*,h^*+p^*+L^Tq^*\rangle=0\nonumber
\end{eqnarray}
This together with \eqref{eq optimality condition GFLasso} proves
\begin{equation}
\lim_{k\rightarrow\infty}V(\beta^k)+\lambda_1\|\beta^k\|_1+\lambda_2\|L\beta^k\|_1=V(\beta^*)+\lambda_1\|\beta^*\|_1+\lambda_2\|L\beta^*\|_1.
\label{eq energy convergence}
\end{equation}

Next, we prove that
\begin{equation}
\lim_{k\rightarrow\infty} \|\beta^k-\beta^*\|_2=0\label{eq uniqueness}
\end{equation}
whenever \eqref{eq generalized fused lasso} has a unique solution.

It is proved by contradiction. Let $\Phi(\beta)=V(\beta)+\lambda_1\|\beta\|_1+\lambda_2\|L\beta\|_1$. Then $\Phi(\beta)$ is a convex, lower continuous function. Since $\beta^*$ is  the unique minimizer, we have $\Phi(\beta)>\Phi(\beta^*)$ for all $\beta\neq \beta^*$. If \eqref{eq uniqueness} does not hold, there exists a subsequence $\beta ^{k_i}$ such that $\|\beta^{k_i}-\beta^*\|>\epsilon$ for some $\epsilon>0$ and for all $i$. Then $\Phi(\beta^{k_i})>\min \{\Phi(\beta):\|\beta-\beta^*\|_2=\epsilon\}$. Indeed, let $\gamma$ be the intersection of the sphere $\{\beta:\|\beta-\beta^*\|_2=\epsilon\}$ and line segment from $\beta^*$ to $\beta^{k_i}$, then there exists a positive number $t\in (0,1)$ such that $\gamma=t\beta^*+(1-t)\beta^{k_i}$. By the convexity of $\Phi$ and the definition of $\beta^*$, we have
\begin{eqnarray}
  \Phi(\beta^{k_i})&>&t\Phi(\beta^*)+(1-t)\Phi(\beta^{k_i})\geq \Phi(t\beta^*+(1-t)\beta^{k_i})\nonumber\\
  &=&\Phi(\gamma)\geq \min \{\Phi(\beta):\|\beta-\beta^*\|_2=\epsilon\}.\nonumber
\end{eqnarray}
Denote $\tilde{\beta}=\arg\min\{\Phi(\beta):\|\beta-\beta^*\|_2=\epsilon\}$ . By applying \eqref{eq energy convergence}, we have
\begin{equation*}
\Phi(\beta^*)=\lim_{i\rightarrow \infty\infty}\Phi(\beta^{k_i})\geq \Phi(\tilde{\beta})>\Phi(\beta^*),
\end{equation*}
which is a contradiction.\eop
\subsection{Convergence analysis of Algorithm \ref{alg SBFLSVM}}
 Now we give the proof of Theorem \ref{theorem convergence analysis SVM}.
The main idea is the same as the one in \cite{CaiOsherShen:MMS:2009}. However, due to the extra bias term $\beta_0$ and the hinge loss, the terms involved in our proof are more complicated.

Since all the subproblems involved in \eqref{eq iteration minimization FLSVM one} are convex, the first order optimality condition gives
\begin{equation}\label{eq iteration minimization FLSVM three}
\begin{cases}
\left(\begin{matrix}
\mu_1I+\mu_2L^TL+\mu_3X^TY^2X&\mu_3X^TYy\cr
\mu_3y^TYX&\mu_3y^Ty
\end{matrix}\right)\left(\begin{matrix}\beta^{k+1}\cr\beta_0^{k+1}\end{matrix}\right)
\cr=\mu_1\left(\begin{matrix}a^k-\mu_1^{-1}u^k\cr 0\end{matrix}\right)+
\mu_2\left(\begin{matrix}L^T\cr 0\end{matrix}\right)(b^k-\mu_2^{-1}v^k)
+\mu_3\left(\begin{matrix}X^TY\cr y^T\end{matrix}\right)(\mathbf{1}-c^k+\mu_3^{-1}w^k)\cr
\lambda_1p^{k+1}-u^k+\mu_1(a^{k+1}-\beta^{k+1})=0,\cr
\lambda_2q^{k+1}-v^k+\mu_2(b^{k+1}-L\beta^{k+1})=0,\cr
\frac{1}{n}s^{k+1}-w^k+\mu_3(c^{k+1}+YX\beta^{k+1}+\beta_0^{k+1}y-\mathbf{1})=0,\cr
u^{k+1}=u^k+\delta_1(\beta^{k+1}-a^{k+1}),\cr
v^{k+1}=v^k+\delta_2(L\beta^{k+1}-b^{k+1}),\cr
w^{k+1}=w^k+\delta_3(\mathbf{1}-YX\beta^{k+1}-\beta_0^{k+1}y-c^{k+1}),
\end{cases}
\end{equation}
where $p^{k+1}\in \partial \|a^{k+1}\|_1,q^{k+1}\in \partial \|b^{k+1}\|_1$ and $s^{k+1}\in \partial \|c^{k+1}\|_1$.
This simple observation will be used in our proof for the convergence of SBFLSVM.

{\bf Proof of Theorem \ref{theorem convergence analysis SVM}.}
Let $(\beta^*,\beta_0^*)$ be an arbitrary minimizer of \eqref{eq fused lasso SVM}. By the first order optimality condition, there exist $p^*, q^*$ and $s^*$ such that
\begin{equation}\label{eq KKT}
\begin{cases}
-\frac{1}{n}X^TYs^*+\lambda_1p^*+\lambda_2L^Tq^*=0,\cr
y^Ts^*=0,
\end{cases}
\end{equation}
where $s_i^*\in\partial(c_i^*)_+$ with $c_i^*=1-y_i(\mathbf{x}_i^T\beta^*+\beta_0^*),i=1,\ldots,n, p^*\in \partial\|\beta^*\|_1$,
$q^*\in\partial\|b^*\|_1$ with $b^*=L\beta^*$. Introducing new variables $a^*=\beta^*,u^*=\lambda_1p^*,v^*=\lambda_2q^*$ and $w^*=\frac{1}{n}s^*$, we can formulate \eqref{eq KKT} as
\begin{equation}\label{eq fixed point}
\begin{cases}
\left(\begin{matrix}
\mu_1I+\mu_2L^TL+\mu_3X^TY^2X&\mu_3X^TYy\cr
\mu_3y^TYX&\mu_3y^Ty
\end{matrix}\right)\left(\begin{matrix}\beta^*\cr\beta_0^*\end{matrix}\right)
\cr=\mu_1\left(\begin{matrix}a^*-\mu_1^{-1}u^*\cr 0\end{matrix}\right)+
\mu_2\left(\begin{matrix}L^T\cr 0\end{matrix}\right)(b^*-\mu_2^{-1}v^*)
+\mu_3\left(\begin{matrix}X^TY\cr y^T\end{matrix}\right)(\mathbf{1}-c^*+\mu_3^{-1}w^*)\cr
\lambda_1p^*-u^*+\mu_1(a^*-\beta^*)=0,\quad \hbox{with}\quad p^*\in \partial \|a^*\|_1\cr
\lambda_2q^*-v^*+\mu_2(b^*-L\beta^*)=0,\quad \hbox{with}\quad q^*\in \partial \|b^*\|_1\cr
\frac{1}{n}s^*-w^*+\mu_3(c^*+YX\beta^*+\beta_0^*y-\mathbf{1})=0,\quad \hbox{with}\quad s^*\in \partial (\sum_{i=1}^n(c_i)_+)\cr
u^*=u^*+\delta_1(\beta^*-a^*),\cr
v^*=v^*+\delta_2(L\beta^*-b^*),\cr
w^*=w^*+\delta_3(\mathbf{1}-YX\beta^*-\beta_0^*y-c^*).\cr
\end{cases}
\end{equation}
Therefore, $\beta^*,\beta_0^*,a^*,b^*,c^*,u^*,v^*,w^*$ is a fixed point of \eqref{eq iteration minimization FLSVM three}. Denote the errors by
\begin{equation*}
\beta_e^k=\beta^k-\beta^*, \beta_{0e}^k=\beta_0^k-\beta_0^*, a_e^k=a^k-a^*, b_e^k=b^k-b^*,
\end{equation*}
$$c_e^k=c^k-c^*,u_e^k=u^k-u^*,v_e^k=v^k-v^*\ \hbox{and} \ w_e^k=w^k-w^*$$
Subtracting the first equation of \eqref{eq iteration minimization FLSVM three} by the first equation of \eqref{eq fixed point}, we obtain
\begin{eqnarray*}
&&\left(\begin{matrix}
\mu_1I+\mu_2L^TL+\mu_3X^TY^2X&\mu_3X^TYy\cr
\mu_3y^TYX&\mu_3y^Ty
\end{matrix}\right)\left(\begin{matrix}\beta_e^{k+1}\cr\beta_{0e}^{k+1}\end{matrix}\right)\nonumber\\
&&=\mu_1\left(\begin{matrix}a_e^k-\mu_1^{-1}u_e^k\cr 0\end{matrix}\right)+
\mu_2\left(\begin{matrix}L^T\cr 0\end{matrix}\right)(b_e^k-\mu_2^{-1}v_e^k)
+\mu_3\left(\begin{matrix}X^TY\cr y^T\end{matrix}\right)(-c_e^k+\mu_3^{-1}w_e^k)
\end{eqnarray*}
Taking the inner product of the left and right hand sides with respect to $((\beta_e^{k+1})^T,\beta_{0e}^{k+1})^T,$ we have
\begin{eqnarray}
&&\langle(\mu_1I+\mu_2L^TL+\mu_3X^TY^2X)\beta_e^{k+1}+\mu_3X^TYy\beta_{0e}^{k+1},
\beta_e^{k+1}\rangle+\mu_3\langle YX\beta_e^{k+1}+y\beta_{0e}^{k+1},y\beta_{0e}^{k+1}\rangle\nonumber\\&=&\mu_1\langle a_e^k,\beta_e^{k+1}\rangle-\langle u_e^k,\beta_e^{k+1}\rangle
+\mu_2\langle b_e^k,L\beta_e^{k+1}\rangle-\langle v_e^k,L\beta_e^{k+1}\rangle\nonumber\\&&-\mu_3\langle c_e^k,YX\beta_e^{k+1}\rangle+\langle w_e^k,YX\beta_e^{k+1}\rangle-\mu_3\langle c_e^k,y\beta_{0e}^{k+1}\rangle+\langle
w_e^k,y\beta_{0e}^{k+1}\rangle.
\label{eq SBFLSVM error one}
\end{eqnarray}
The same manipulations applied to the second (third, fourth) of equation \eqref{eq iteration minimization FLSVM three} and the second (third, fourth) of equation \eqref{eq fixed point} lead to
\begin{equation}\label{eq SBFLSVM error two}
\lambda_1\langle p^{k+1}-p^*,a^{k+1}-a^*\rangle+\mu_1\|a_e^{k+1}\|^2_2-\langle u_e^k,a_e^{k+1}\rangle-\mu_1\langle \beta_e^{k+1},a_e^{k+1}\rangle=0.
\end{equation}
\begin{equation}\label{eq SBFLSVM error three}
\lambda_2\langle q^{k+1}-q^*,b^{k+1}-b^*\rangle+\mu_2\|b_e^{k+1}\|_2^2-\langle v_e^k,b_e^{k+1}\rangle-\mu_2\langle L\beta_e^{k+1},b_e^{k+1}\rangle=0.
\end{equation}
\begin{equation}\label{eq SBFLSVM error four}
\frac{1}{n}\langle s^{k+1}-s^*,c^{k+1}-c^*\rangle
+\mu_3\|c_e^{k+1}\|_2^2-\langle w_e^k,c_e^{k+1}\rangle+\mu_3\langle YX\beta_e^{k+1}+\beta_{0e}^{k+1}y,c_e^{k+1}\rangle=0.
\end{equation}
By summing equations \eqref{eq SBFLSVM error one}, \eqref{eq SBFLSVM error two}, \eqref{eq SBFLSVM error three} and \eqref{eq SBFLSVM error four}, we get
\begin{eqnarray}
&&\quad\lambda_1\langle p^{k+1}-p^*,a^{k+1}-a^*\rangle+\lambda_2\langle q^{k+1}-q^*,b^{k+1}-b^*\rangle+\frac{1}{n}\langle s^{k+1}-s^*,c^{k+1}-c^*\rangle\nonumber\\
&&+\mu_1(\|\beta_e^{k+1}\|_2^2
-\langle\beta_e^{k+1},a_e^{k+1}+a_e^k\rangle+\|a_e^{k+1}\|_2^2)\nonumber\\&&+\mu_2(\|L\beta_e^{k+1}\|_2^2-\langle b_e^{k+1}+b_e^k,L\beta_e^{k+1}\rangle+\|b_e^{k+1}\|_2^2)\nonumber\\
&&+\mu_3(\|YX\beta_e^{k+1}+y\beta_{0e}^{k+1}\|_2^2+
\mu_3\langle YX\beta_e^{k+1}+y\beta_{0e}^{k+1},c_e^{k+1}+c_e^k\rangle+\|c_e^{k+1}\|_2^2)\nonumber\\
&&+\langle u_e^k,\beta_e^{k+1}-a_e^{k+1}\rangle+\langle v_e^k,L\beta_e^{k+1}-b_e^{k+1}\rangle
-\langle w_e^k,YX\beta_e^{k+1}+y\beta_{0e}^{k+1}+c_e^{k+1}\rangle=0.\nonumber\\\label{eq summation}
\end{eqnarray}
Furthermore, subtracting the fifth equation of \eqref{eq iteration minimization FLSVM three} by the fifth equation of \eqref{eq fixed point}, we have
\begin{equation*}
u_e^{k+1}=u_e^{k}+\delta_1(\beta_e^{k+1}-a_e^{k+1}).
\end{equation*}
which leads to
\begin{equation}\label{eq SBFSVM error five}
\langle u_e^k,\beta_e^{k+1}-a_e^{k+1}\rangle=\frac{1}{2\delta_1}
(\|u_e^{k+1}\|_2^2-\|u_e^k\|_2^2)-\frac{\delta_1}{2}\|\beta_e^{k+1}-a_e^{k+1}\|_2^2.
\end{equation}
Similarly, we can get
\begin{equation}\label{eq SBFLSVM error six}
\langle v_e^k, L\beta_e^{k+1}-b_e^{k+1}\rangle=\frac{1}{2\delta_2}(\|v_e^{k+1}\|_2^2
-\|v_e^k\|_2^2)-\frac{\delta_2}{2}\|L\beta_e^{k+1}-b_e^{k+1}\|_2^2
\end{equation}
and
\begin{eqnarray}\label{eq SBFLSVM seven}
&&\langle w_e^k,YX\beta_e^{k+1}+\beta_{0e}^{k+1}y+c_e^{k+1}\rangle\nonumber\\&=&-\frac{1}{2\delta_3}
(\|w_e^{k+1}\|_2^2-\|w_e^k\|_2^2)+\frac{\delta_3}{2}\|YX\beta_e^{k+1}+\beta_{0e}^{k+1}y+c_e^{k+1}\|_2^2.
\end{eqnarray}
Substituting \eqref{eq SBFSVM error five},\eqref{eq SBFLSVM error six} and \eqref{eq SBFLSVM seven} into \eqref{eq summation} yields
\begin{eqnarray}
&&\frac{1}{2\delta_1}(\|u_e^k\|_2^2-\|u_e^{k+1}\|_2^2)+\frac{1}{2\delta_2}
(\|v_e^k\|_2^2-\|v_e^{k+1}\|_2^2)+\frac{1}{2\delta_3}(\|w_e^k\|_2^2-\|w_e^{k+1}\|_2^2)\nonumber\\
&=&\lambda_1\langle p^{k+1}-p^*,a^{k+1}-a^*\rangle+\lambda_2\langle q^{k+1}-q^*,b^{k+1}-b^*\rangle+\frac{1}{n}\langle s^{k+1}-s^*,c^{k+1}-c^*\rangle\nonumber\\
&&+\mu_1\bigg(\|\beta_e^{k+1}\|_2^2
-\langle\beta_e^{k+1},a_e^{k+1}+a_e^k\rangle+\|a_e^{k+1}\|_2^2-\frac{\delta_1}{2\mu_1}
\|\beta_e^{k+1}-a_e^{k+1}\|_2^2\bigg)+\mu_2\bigg(\|L\beta_e^{k+1}\|_2^2\nonumber\\&&-\langle L\beta_e^{k+1},b_e^{k+1}+b_e^k\rangle+\|b_e^{k+1}\|_2^2-\frac{\delta_2}{2\mu_2}
\|L\beta_e^{k+1}-b_e^{k+1}\|_2^2\bigg)+\mu_3\bigg(\|YX\beta_e^{k+1}+y\beta_{0e}^{k+1}\|_2^2\nonumber\\
&&+
\langle YX\beta_e^{k+1}+y\beta_{0e}^{k+1},c_e^{k+1}+c_e^k\rangle+\|c_e^{k+1}\|_2^2-
\frac{\delta_3}{2\mu_3}\|YX\beta_e^{k+1}+\beta_{0e}^{k+1}y+c_e^{k+1}\|_2^2\bigg).\nonumber\\
\label{eq summation one}
\end{eqnarray}

Using the elementary equality \eqref{eq elementary equation}, \eqref{eq summation one} can be transformed to
\begin{eqnarray}
&&\frac{1}{2\delta_1}(\|u_e^k\|_2^2-\|u_e^{k+1}\|_2^2)+\frac{1}{2\delta_2}
(\|v_e^k\|_2^2-\|v_e^{k+1}\|_2^2)+\frac{1}{2\delta_3}(\|w_e^k\|_2^2-\|w_e^{k+1}\|_2^2)\nonumber\\
&=&\lambda_1\langle p^{k+1}-p^*,a^{k+1}-a^*\rangle+\lambda_2\langle q^{k+1}-q^*,b^{k+1}-b^*\rangle+\frac{1}{n}\langle s^{k+1}-s^*,c^{k+1}-c^*\rangle\nonumber\\
&&+\frac{\mu_1}{2}\bigg(\|\beta_e^{k+1}-a_e^k\|_2^2
+\|a_e^{k+1}\|_2^2-\|a_e^k\|_2^2+\frac{\mu_1-\delta_1}{\mu_1}
\|\beta_e^{k+1}-a_e^{k+1}\|_2^2\bigg)\nonumber\\&&+\frac{\mu_2}{2}\bigg(\|L\beta_e^{k+1}-b_e^k\|_2^2 +\|b_e^{k+1}\|_2^2-\|b_e^k\|_2^2+\frac{\mu_2-\delta_2}{\mu_2}
\|L\beta_e^{k+1}-b_e^{k+1}\|_2^2\bigg)\nonumber\\&&+\frac{\mu_3}{2}\bigg(\|YX\beta_e^{k+1}+y\beta_{0e}^{k+1}+c_e^k\|_2^2
+\|c_e^{k+1}\|_2^2-\|c_e^k\|_2^2
\nonumber\\&&+\frac{\mu_3-\delta_3}{\mu_3}\|YX\beta_e^{k+1}+\beta_{0e}^{k+1}y+c_e^{k+1}\|_2^2\bigg).\nonumber\\
\label{eq summation two}
\end{eqnarray}
Summing the above equation from $k=0$ to $k=K$ yields
\begin{eqnarray}
&&\frac{1}{2\delta_1}(\|u_e^0\|_2^2-\|u_e^{K+1}\|_2^2)+\frac{1}{2\delta_2}
(\|v_e^0\|_2^2-\|v_e^{K+1}\|_2^2)+\frac{1}{2\delta_3}(\|w_e^0\|_2^2-\|w_e^{K+1}\|_2^2)\nonumber\\
&&+\frac{\mu_1}{2}(\|a_e^0\|_2^2-\|a_e^{K+1}\|_2^2)+\frac{\mu_2}{2}(\|b_e^0\|_2^2-\|b_e^{K+1}\|_2^2)
+\frac{\mu_3}{2}(\|c_e^0\|_2^2-\|c_e^{K+1}\|_2^2)\nonumber\\
&=&\lambda_1\sum_{k=0}^K\langle p^{k+1}-p^*,a^{k+1}-a^*\rangle+\lambda_2\sum_{k=0}^K\langle q^{k+1}-q^*,b^{k+1}-b^*\rangle\nonumber\\
&&+\frac{1}{n}\sum_{k=0}^K\langle s^{k+1}-s^*,c^{k+1}-c^*\rangle+\frac{\mu_1}{2}\sum_{k=0}^K\|\beta_e^{k+1}-a_e^k\|_2^2+\frac{\mu_1-\delta_1}{2}
\sum_{k=0}^K\|\beta_e^{k+1}-a_e^{k+1}\|_2^2\nonumber\\
&&+\frac{\mu_2}{2}\sum_{k=0}^K\|L\beta_e^{k+1}-b_e^k\|_2^2+\frac{\mu_2-\delta_2}{2}\sum_{k=0}^K\|L\beta_e^{k+1}-b_e^{k+1}\|_2^2\nonumber\\&&
+\frac{\mu_3}{2}\sum_{k=0}^K\|YX\beta_e^{k+1}+y\beta_{0e}^{k+1}+c_e^k\|_2^2+
\frac{\mu_3-\delta_3}{2}\sum_{k=0}^K\|YX\beta_e^{k+1}+\beta_{0e}^{k+1}y+c_e^{k+1}\|_2^2.\nonumber\\\label{eq summation three}
\end{eqnarray}
The fact $p^{k+1}\in \partial \|a^{k+1}\|_1, p^*\in \partial \|a^*\|$ and $\|\cdot\|_1$ is convex implies $\langle p^{k+1}-p^*, a^{k+1}-a^*\rangle\geq 0$. Similarly,
$\langle q^{k+1}-q^*,b^{k+1}-b^*\rangle\geq 0,\langle s^{k+1}-s^*,c^{k+1}-c^*\rangle\geq 0$. Therefore,
all terms involved in \eqref{eq summation three} are nonnegative. Now we can cheat each term in the right hand side of \eqref{eq summation three} separately by the same argument as the proof of Theorem \ref{theorem convergence analysis} and get the convergence result \eqref{eq SBFLSVM convergence energy}. The proof of \eqref{eq SBFLSVM convergence beta} can also follows the same line as the one of Theorem \ref{theorem convergence analysis}, we omit the details here.

\subsection{Updates in SBFLSVM \label{alg update SBFLSVM}}
\subsubsection{Update of $\beta$}
Due to the extra bias term of $\beta_0^{k+1}$, we need to solve the following linear system which is slightly different from \eqref{eq linear equation SBFLasso}.
\begin{eqnarray}
&&\left(\begin{matrix}
\mu_1I+\mu_2L^TL+\mu_3X^TY^2X&\mu_3X^TYy\cr
\mu_3y^TYX&\mu_3y^Ty
\end{matrix}\right)\left(\begin{matrix}\beta^{k+1}\cr\beta_0^{k+1}\end{matrix}\right)\nonumber\\
&=&\mu_1\left(\begin{matrix}a^k-\mu_1^{-1}u^k\cr 0\end{matrix}\right)+
\mu_2\left(\begin{matrix}L^T\cr 0\end{matrix}\right)(b^k-\mu_2^{-1}v^k)
+\mu_3\left(\begin{matrix}X^TY\cr y^T\end{matrix}\right)(\mathbf{1}-c^k+\mu_3^{-1}w^k)\nonumber\\\label{eq linear equation FLSVM}
\end{eqnarray}
Fortunately, this linear system can also be solved by PCG efficiently. Note that
\begin{eqnarray*}
&&\left(\begin{matrix}
\mu_1I+\mu_2L^TL+\mu_3X^TY^2X&\mu_3X^TYy\cr
\mu_3y^TYX&\mu_3y^Ty
\end{matrix}\right)\\&=&\left(\begin{matrix}\mu_1I+\mu_2L^TL&0\cr0&\mu_3y^Ty\end{matrix}\right)
+\mu_3\left(\begin{matrix}X^TY\cr y^T\end{matrix}\right)(YX,y)-\left(\begin{matrix}0&0\cr 0&\mu_3y^Ty \end{matrix}
\right).
\end{eqnarray*}
It is easy to see that $\left(\begin{matrix}\mu_1I+\mu_2L^TL&0\cr0&\mu_3y^Ty\end{matrix}\right)$ is still a tridiagonal matrix and $$\mu_3\left(\begin{matrix}X^TY\cr y^T\end{matrix}\right)(YX,y)-\left(\begin{matrix}0&0\cr 0&\mu_3y^Ty \end{matrix}
\right)$$ is a low rank matrix. So PCG is still a good solver for the linear system \eqref{eq linear equation FLSVM}.

\subsubsection{Proof of Proposition \ref{prop shrinkage operator}}

\begin{proof}
The energy function $\lambda x_++\frac{1}{2}\|x-w\|_2^2$ is strongly convex, hence has a unique minimizer. Therefore, by the subdifferential calculus (c.f. \cite{HL:BOOK:93}), $s_\lambda$ is the unique solution of the following equation with unknown $w$
\begin{equation}\label{eq subdifferential equation}
0\in \lambda\partial(x_+)+x-w,
\end{equation}
where $\partial(x_+)=\{p\in \mathbb{R}:y_+-x_+-(y-x)p\geq 0,\forall y\in \mathbb{R}\}$
is the subdifferential of the function $x_+.$ If $x\neq 0$, then $x_+$ is differentiable, and
its subdifferential contains only its gradient. If $x=0$, then $\partial(x_+)=\{p\in\mathbb{R}:y_+-yp\geq0,\forall y\in \mathbb{R}\}.$
One can check that $\partial(x_+)=\{p:0\leq p\leq 1\}$ for this case. Indeed, for any $p\in [0,1],$ $yp\leq y_+$ by using the definition of $y_+$. On the other hand, if there exists a number $p\in (-\infty,0)\cup(1,+\infty)$ and $p\in \partial
(x_+)$, then  we can easily get a contraction. Actually, the fact $p\in (-\infty,0)\cup(1,+\infty)$ implies $p^2>p_+$. On the other hand, since $\partial(x_+)=\{p\in\mathbb{R}:y_+-yp\geq0,\forall y\in \mathbb{R}\}$ for $x=0$, we have $p^2<p_+$  by letting $y=p$. In summary,
\begin{equation}\label{eq subdifferential of hinge loss}
\partial(x_+)=\left\{\begin{array}{ll}
1,&x>0,\\\{p:p\in[0,1]\},&x=0,\\
0,&x<0.
\end{array}
\right.
\end{equation}
With \eqref{eq subdifferential equation} and \eqref{eq subdifferential of hinge loss}, we can get the desired result.
\end{proof}

\end{document}